\documentclass[aps, prd, amsmath, floats, floatfix, twocolumn, nofootinbib, superscriptaddress]{revtex4-1}
\usepackage[pdftex]{graphicx}
\usepackage{color}
\usepackage{latexsym}
\usepackage[colorlinks]{hyperref}

\newcommand{\be}{\begin{eqnarray}}
\newcommand{\ee}{\end{eqnarray}}

\begin{document}

\title{Probing the near-horizon region of Cygnus~X-1 with \textsl{Suzaku} and \textsl{NuSTAR}}

\author{Zuobin~Zhang}
\affiliation{Center for Field Theory and Particle Physics and Department of Physics, Fudan University, 200438 Shanghai, China}

\author{Honghui~Liu}
\affiliation{Center for Field Theory and Particle Physics and Department of Physics, Fudan University, 200438 Shanghai, China}

\author{Askar~B.~Abdikamalov}
\affiliation{Center for Field Theory and Particle Physics and Department of Physics, Fudan University, 200438 Shanghai, China}
\affiliation{Ulugh Beg Astronomical Institute, Tashkent 100052, Uzbekistan}

\author{Dimitry~Ayzenberg}
\affiliation{Theoretical Astrophysics, Eberhard-Karls Universit\"at T\"ubingen, D-72076 T\"ubingen, Germany}

\author{Cosimo~Bambi}
\email[Corresponding author: ]{bambi@fudan.edu.cn}
\affiliation{Center for Field Theory and Particle Physics and Department of Physics, Fudan University, 200438 Shanghai, China}

\author{Menglei~Zhou}
\affiliation{Institut f\"ur Astronomie und Astrophysik, Eberhard-Karls Universit\"at T\"ubingen, D-72076 T\"ubingen, Germany}

\begin{abstract}
Astrophysical black holes are ideal laboratories for testing Einstein's theory of general relativity in the strong field regime. In this manuscript, we present an analysis of \textsl{Suzaku} and \textsl{NuSTAR} spectra of the black hole binary Cygnus~X-1 using {\tt relxill\_nk}. Unlike our previous study on Cygnus~X-1 with \textsl{NuSTAR} data, here we are able to constrain the Johannsen deformation parameter $\alpha_{13}$. However, despite the high energy resolution near the iron line provided by \textsl{Suzaku}, our constraints on the Kerr metric from Cygnus~X-1 are not very stringent in comparison with those that have been obtained from other sources, confirming that Cygnus~X-1 is quite a complicated source.
\end{abstract}

\maketitle


\section{Introduction}

Einstein's theory of general relativity was proposed over a century ago and so far has successfully passed a large number of observational tests. Since the 1960s, there have been significant efforts to test general relativity in the so-called weak field regime with experiments in the Solar System and observations of binary pulsars~\cite{Will:2014kxa}. In the past 20~years, there has been an increasing interest in testing general relativity on large scales with cosmological tests, which are mainly motivated by the problems of dark matter and dark energy~\cite{ Jain:2010ka,Koyama:2015vza,Ferreira:2019xrr}. Thanks to new observational facilities, the past 5~years have seen a tremendous progress on tests of general relativity in the so-called strong field regime, and we are now starting testing the predictions of Einstein's gravity in the strong gravitational field of black holes~\cite{TheLIGOScientific:2016src,Yunes:2016jcc,RMP,PRL,MCG,Bambi:2019tjh,Tripathi:2020qco,Abdikamalov:2020oci,Psaltis:2020lvx,Tripathi:2020dni,Abbott:2020jks,Carson:2020rea}.

The spacetime metric around astrophysical black holes is thought to be approximated well by the Kerr solution~\cite{kerr}, since the possible electric charge of the object or the gravitational field produced by the accretion disk or nearby stars is normally completely negligible in the near-horizon region~\cite{k1,k2,k3}. However, a number of authors have pointed out that new physics may produce macroscopic deviations from the Kerr solution, because general relativity may not be the correct theory of gravity, there may be large quantum gravity effects, or because of the presence of exotic matter fields; see, e.g. Refs.~\cite{Dvali:2011aa,Herdeiro:2014goa,Giddings:2014ova,Ayzenberg:2014aka,Giddings:2016btb}.

As of now, tests of general relativity in the strong gravity regime using black holes are possible by studying the gravitational wave signal from the coalescence of stellar-mass black holes (e.g.~\cite{TheLIGOScientific:2016src,Yunes:2016jcc,Abbott:2020jks,Carson:2020rea}), by studying the properties of the X-ray radiation emitted from the inner part of the accretion disk (e.g.~\cite{PRL,MCG,Tripathi:2020qco,Abdikamalov:2020oci,Tripathi:2020dni}), or by studying the image of the supermassive black holes in the galaxy M87 (e.g.~\cite{Bambi:2019tjh,Psaltis:2020lvx}). In the past few years, our group has mainly worked on X-ray tests by developing the relativistic reflection model {\tt relxill\_nk}~\cite{noi1,noi2}, which is an extension of the {\tt relxill} package~\cite{j1,j2,j3} to non-Kerr spacetimes. In addition to the parameters present in {\tt relxill}, {\tt relxill\_nk} has some ``deformation parameters'' to quantify possible deviations from the Kerr geometry, which is recovered when all deformation parameters vanish. The reflection spectrum of the accretion disk is calculated without assuming that these deformation parameters vanish. From the comparison of the theoretical model with the observational data, it is possible to estimate the value of the deformation parameters and check whether the observations are consistent with the hypothesis that the spacetime metric around astrophysical black holes is described by the Kerr solution as expected in the standard framework.
Like most astrophysical measurements, the accuracy of the estimate of the deformation parameters depends on the capability of the theoretical model to properly describe the astrophysical system and it is thus very important to select the sources and the observations that are expected to match better the characteristics of the theoretical model in order to limit undesirable systematic uncertainties.

The origin of the reflection spectrum of the disk can be understood as follows. We consider a black hole accreting from a geometrically thin and optically thick disk. The thermal spectrum of the disk is peaked in the soft X-ray band in the case of stellar-mass black holes and in the optical/UV band in the case of supermassive black holes. Thermal photons from the disk inverse Compton scatter off free electrons in the corona, which is some hotter ($\sim 100$~keV) cloud near the black hole and the inner part of the accretion disk, even if its morphology is not yet well understood. The Comptonized photons have a power-law spectrum with a high-energy cut-off and can illuminate the accretion disk, producing a reflection component. The latter is characterized by some fluorescent emission lines below 10~keV, notably the iron K$\alpha$ complex, and by a Compton hump peaked at 20-30~keV. The analysis of the reflection features can be used to study the morphology of the accreting matter, measure black hole spins, and test Einstein's theory of general relativity in the strong field regime~\cite{csr,Bambi:2020jpe}.

The analysis of relativistic reflection features requires both a good energy resolution near the iron line and data over a broad energy band. The good energy resolution near the iron line is useful because the shape of the iron K$\alpha$ line is the most informative part of the spectrum concerning the motion of the gas in the very strong gravity region around the black hole. The broad energy band helps to select the correct astrophysical model and to constrain better some disk-related parameters that, in turn, help to measure better the parameters of the spacetime metric.

In Ref.~\cite{Liu:2019vqh}, we analyzed with {\tt relxill\_nk} two \textsl{NuSTAR} observations of Cygnus~X-1, respectively in 2012 and 2014, when the source was in the soft state. We found it very challenging to test the Kerr metric from those spectra, in the sense that our measurements of the deformation parameters could easily change by changing some assumptions like the emissivity profile model for the accretion disk. Similar problems were met with the analysis of a \textsl{NuSTAR} observation of GRS~1915+105~\cite{yxz} and of \textsl{RXTE} data of GX~339--4~\cite{Wang-Ji:2018ssh}. On the contrary, very promising results have been found from the analysis of \textsl{Suzaku} and \textsl{XMM-Newton}+\textsl{NuSTAR} observations of other sources~\cite{MCG,bare,yxz2}, suggesting the possibility that for testing the Kerr metric it is strictly necessary to have a good energy resolution near the iron line. In the present paper, we thus analyze a 2009 \textsl{Suzaku} observation of Cygnus~X-1 in the hard state and a simultaneous observation \textsl{Suzaku}+\textsl{NuSTAR} in 2012 when Cygnus~X-1 was in the soft state.

Our manuscript is organized as follows. In Section~\ref{s-red}, we present the \textsl{Suzaku} and \textsl{NuSTAR} observations analyzed in our work and we describe their data reduction. In Section~\ref{s-ana-h}, we report the analysis of the 2009 \textsl{Suzaku} observation, when the source was in the hard state. Section~\ref{s-ana-s} is devoted to the analysis of the simultaneous observations \textsl{Suzaku}+\textsl{NuSTAR} in 2012, when Cygnus~X-1 was in the soft state. We discuss our results in Section~\ref{s-d-c}.


\begin{table*}
\centering
\vspace{0.5cm}
\begin{tabular}{ccccccc}
\hline\hline
Epoch & \hspace{0.5cm} State \hspace{0.5cm}  & \hspace{0.5cm} Mission \hspace{0.5cm}  & \hspace{0.5cm} Obs.~ID \hspace{0.5cm} & \hspace{0.5cm} Instrument(s) \hspace{0.5cm} & \hspace{0.5cm} Start date \hspace{0.5cm} & \hspace{0.5cm} Exposure~(ks) \hspace{0.5cm} \\
\hline\hline
1 & Hard & \textsl{Suzaku} &404075020 &XIS0+3 & 2009-04-08 & 10.7 \\
  & & & & HXD/PIN& & 12.2 \\
  & & & & HXD/GSO& &  12.2 \\
\hline
2 & Soft & \textsl{Suzaku} &407072010 &XIS0+1 & 2012-10-31 & 1.9 \\
  & & & & HXD/PIN&  & 30.1 \\
  & & & & HXD/GSO&  &  27.9 \\
  & & \textsl{NuSTAR} & 30001011002 &FPMA+FPMB & 2012-10-31 & 11.0 \\
  & & & 30001011003 &FPMA+FPMB & & 5.7 \\
  & & & 10014001001 &FPMA+FPMB & & 4.6 \\
\hline\hline
\end{tabular}
\caption{Summary of the observations analyzed in the present work. \label{t-obs}}
\end{table*}

\section{Observations and data reduction \label{s-red}}

Cygnus~X-1 was discovered in 1964 during a rocket flight and is one of the brightest X-ray sources in the sky. It is a binary system with a black hole with a mass of $14.8 \pm 1.0$~$M_{\odot}$~\cite{Orosz:2011np} and a type O9.7Iab supergiant companion star. The strong wind from the companion star provides a sufficiently high mass transfer to the black hole at any time, so Cygnus~X-1 is a persistent X-ray source. The distance of this binary system from us is $1.86^{+0.12}_{-0.11}$~kpc~\cite{Reid:2011nn}.

In the present work, we consider the \textsl{Suzaku}~\cite{suzaku} and \textsl{NuSTAR}~\cite{nustar} observations listed in Tab.~\ref{t-obs}. The first observation (epoch~1) was in 2009 by \textsl{Suzaku}, when the source was in a hard state. The second observation (epoch~2) refers to a simultaneous observation of \textsl{Suzaku} and \textsl{NuSTAR} in 2012, when the source was in a soft state. Spectra from the data are created by using the tools from the HEASOFT~6.26 package: version 20181023 of Suzaku Calibration Database (CALDB) for the \textsl{Suzaku} data and version 20180419 of NuSTAR Calibration Database (CALDB) for the \textsl{NuSTAR} data.

\subsection{\textsl{Suzaku} data in the hard state}

The \textsl{Suzaku} satellite was equipped with three detectors (XISs, HXD/PIN, and HXD/GSO), through which \textsl{Suzaku} could cover the 0.3-600~keV band. During the observation of the epoch~1, only XIS0, XIS3, PIN, and GSO detectors were operational. The XIS data are reduced following the standard procedure, which includes {\tt xispi} for energy scale reprocessing and {\tt xisrepro} for data screening. For each event file, we use the {\tt aeattcor2} tool to correct the data for large spacecraft wobble endemic to \textsl{Suzaku}.

An issue in the \textsl{Suzaku} data of Cygnus~X-1 is the presence of photon pile-up, as a consequence of the brightness of the source. To minimize the pile-up effect, we use the FTOOL {\tt pileest} to estimate the pile-up degree and thus exclude the most heavily piled-up region. Finally, we extract the spectra from a square box, 240~pixels on each side, with an inner exclusion circle with a radius of 30~pixels, which is centered on the source coordinates. This effectively reduced the pile-up on the extracted region to $5\%$ or less. The redistribution matrix files and ancillary response files are constructed by the tool {\tt xisrmfgen} and {\tt xissimarfgen}, respectively. Since the XIS0 and XIS3 CCDs are both front-illuminated chips and have similar response function, we use {\tt addascaspec} to combine their spectra, backgrounds, and response functions. We then grouped the spectra to have at least 10~counts per bin as done in~\cite{Miller:2012zj}.

The HXD/PIN and HXD/GSO spectra are generated and reduced using the FTOOL {\tt hxdpinxbpi} and {\tt hxdgsoxbpi}, respectively. The corresponding background and response (including the ``correction arf'') files are downloaded from the {\tt pinxb\_ver2.0\_tuned} and {\tt gsonxb\_ver2.0} directory at HEASARC. Eventually, we analyze the data in the 0.8-1.7~keV and 2.5-9.0~keV bands for the XIS spectra, the 12-70~keV band for PIN, and 70-500~keV band for GSO.

\subsection{\textsl{Suzaku} data in the soft state}

For the \textsl{Suzaku} observation 407072010, we proceed in a similar way. Since the source was in the soft state, the data are more heavily affected by pile-up than those of the 2009 observation. The spectra are extracted from a rectangular region, approximately $320\times220$~pixels, with two inner exclusion rectangles with the size of $130\times45$~pixels, centered on the source coordinates. This effectively reduced the pile-up on the extracted region to 5\% or less. Because the XIS0 CCD has front-illuminated chips while XIS1 has back-illuminated chips, the spectra were separately grouped using the tool {\tt grppha} so that each bin have at least 50~counts. For the HXD/PIN and HXD/GSO data, we still used, respectively, the FTOOL {\tt hxdpinxbpi} and {\tt hxdgsoxbpi} to extract the spectra. We use data in the 1.2-1.7~keV and 2.5-9.0~keV band from XIS, 15-68~keV band from PIN, and 50-296~keV band from GSO.

\subsection{\textsl{NuSTAR} data in the soft state}

We reduce the data from instruments Focal Plane Modules A and B (FPMA and FPMB). We separately reduce the data of the three observations and then use {\tt addascaspec} to combine the spectra for each FPM instrument. We run {\tt nuproduct} to extract light curves, spectra, and response files. The source is extracted from a circular region centered on Cygnus~X-1 with a radius of 150''. The background region is a circle with the same size taken far from the source region to avoid any contribution from the source. The spectra are then grouped to have a minimum count of 50~photons per bin.


\begin{table*}
\centering
\vspace{0.5cm}
\begin{tabular}{ccc}
\hline\hline
\hspace{0.5cm} Model name \hspace{0.5cm} & XSPEC model & \hspace{0.5cm} $q_{\rm out}$ \hspace{0.5cm} \\
\hline\hline
1A & {\tt tbabs$\times$xstar$\times$(diskbb+cutoffpl+gauss+relxill\_nk)} & 3 \\
1B &  & free \\
\hline
2A & {\tt tbabs$\times$xstar$\times$(diskpbb+cutoffpl+gauss+relxill\_nk)} & 3 \\
2B &  & free \\
\hline
3A & {\tt tbabs$\times$xstar$\times$(simplcutx$\times$diskpbb+gauss+relxillCp\_nk)} & 3 \\
3B &  & free \\
\hline\hline
\end{tabular}
\caption{Summary of the models used to fit the data of epoch~1. \label{t-mod}}
\end{table*}

\section{Analysis of epoch~1 \label{s-ana-h}}

Here and in the next section, we use XSPEC v12.10.1s for the spectral analysis~\cite{xspec}. The Galactic absorption is fitted with {\tt tbabs}~\cite{wilms-2000}. The relativistic reflection component is fitted with {\tt relxill\_nk} v1.3.3~\cite{noi1,noi2}. The spacetime metric around the black hole is described by the Johannsen metric with the deformation parameter $\alpha_{13}$, while all other deformation parameters vanish~\cite{tj}.

\begin{figure}[htbp]
	\centering
	\includegraphics[scale=0.32]{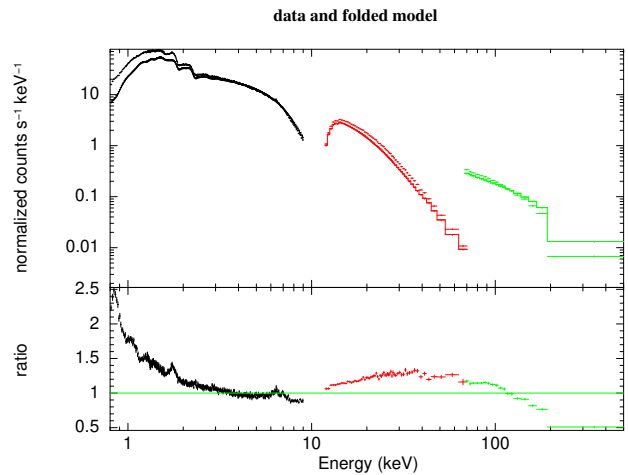}
	\vspace{-0.6cm}
	\caption{The upper quadrant shows the spectrum of Cygnus~X-1 in epoch~1 as seen by XIS0+XIS3 (black), PIN (red) and GSO (green). The spectrum is fitted with an absorbed power-law ignoring the data below 3~keV, the 5-7~keV band, and the 15-45~keV band. The data to the best-fit model ratio is shown in the lower quadrant. The data have been rebinned for visual clarity. 
	\label{f-c1}}	 
\end{figure}

To start, we fit the 2009 \textsl{Suzaku} data with a simple power law. The resulting spectrum and data-to-model ratio are shown in Fig.~\ref{f-c1}. We clearly see an excess of photons below 2~keV, a small feature around 6-7~keV, and a bump in the HXD data. The excess of photons at low energies suggests the presence of the thermal component from the accretion disk. The small feature at 6-7~keV and the bump can be naturally interpreted as an iron line and a Compton hump from the reflection spectrum.


The models used to fit the 2009 \textsl{Suzaku} data are summarized in Tab.~\ref{t-mod}. We use {\tt diskbb} to describe the thermal spectrum of the accretion disk~\cite{diskbb}. We use {\tt relxill\_nk} to fit the reflection spectrum setting the reflection fraction to $-1$ and using {\tt cutoffpl} to describe the Comptonized photons from the corona. We find that we can get a better fit adding a Gaussian emission line. Eventually the XSPEC model is (Model~1):

\vspace{0.2cm}

\noindent {\tt tbabs$\times$xstar$\times$(diskbb+cutoffpl+gauss+relxill\_nk)},

\vspace{0.2cm}

\noindent where {\tt xstar} describes an ionized plasma in the wind from the companion star. A multiplicative constant is added to combine the data from different instruments. In {\tt relxill\_nk}, we use a broken power-law to describe the emissivity profile of the accretion disk and the outer emissivity index $q_{\rm out}$ is either frozen to 3 (Model~1A) or left free in the fit (Model~1B). Note that here and in what follows, for every model we fit the data first assuming the Kerr metric and setting $\alpha_{13} = 0$ and then leaving $\alpha_{13}$ free in the fit.

\begin{figure*}[t]
	\begin{center}
		\includegraphics[scale=0.54]{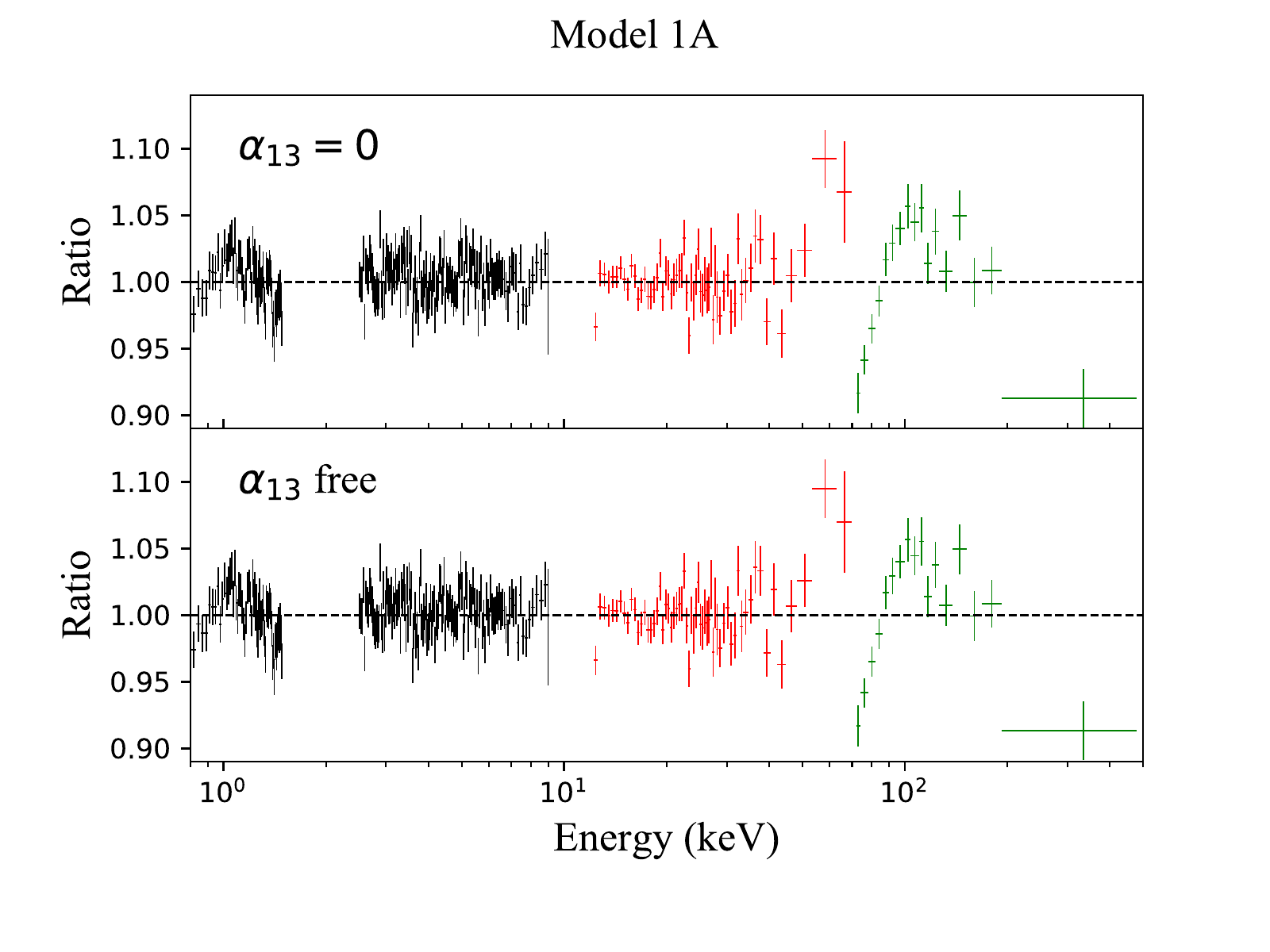}
		\includegraphics[scale=0.54]{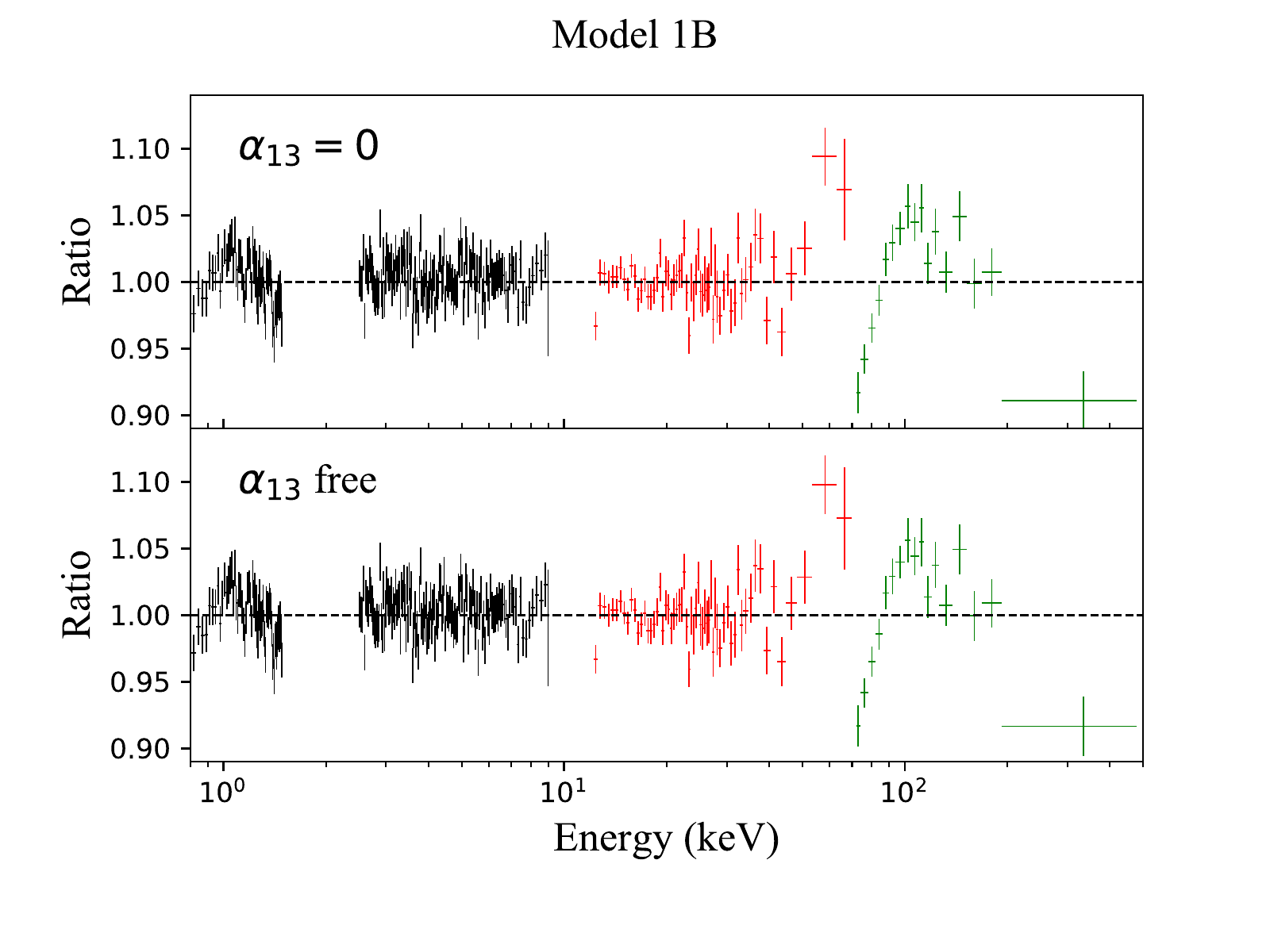} \\ \vspace{-0.3cm}
		\includegraphics[scale=0.54]{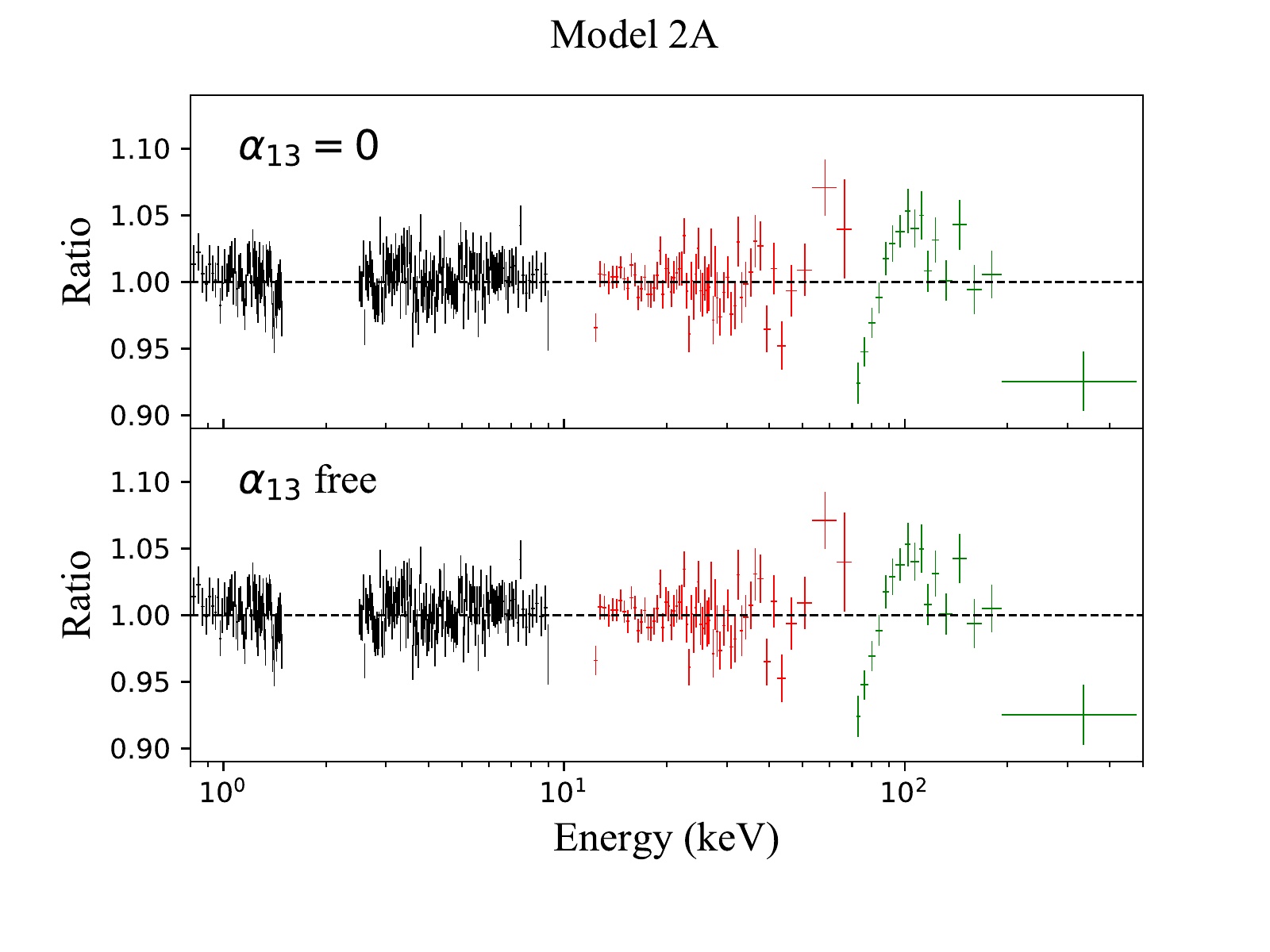}
		\includegraphics[scale=0.54]{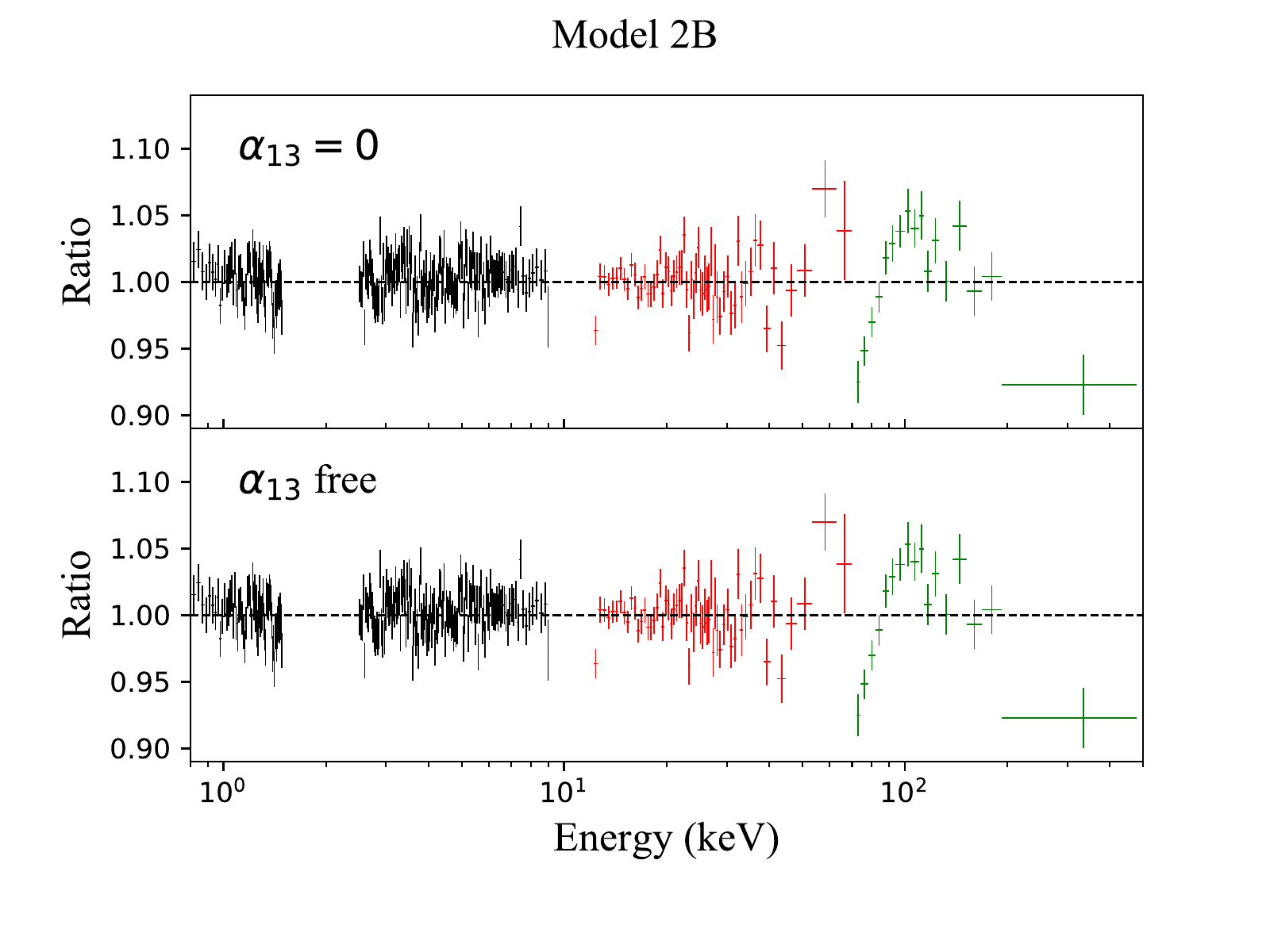} \\ \vspace{-0.3cm}
		\includegraphics[scale=0.54]{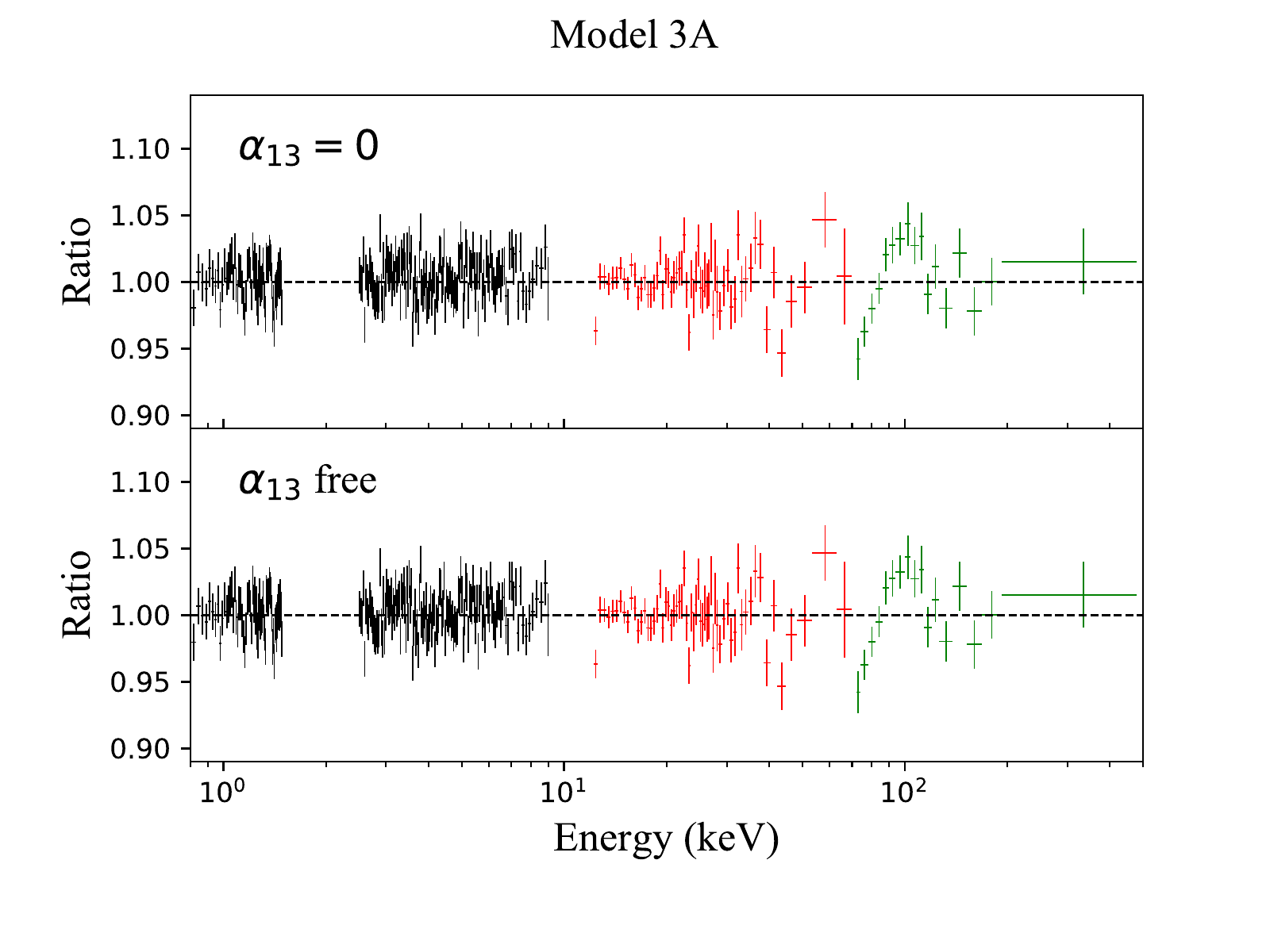}
		\includegraphics[scale=0.54]{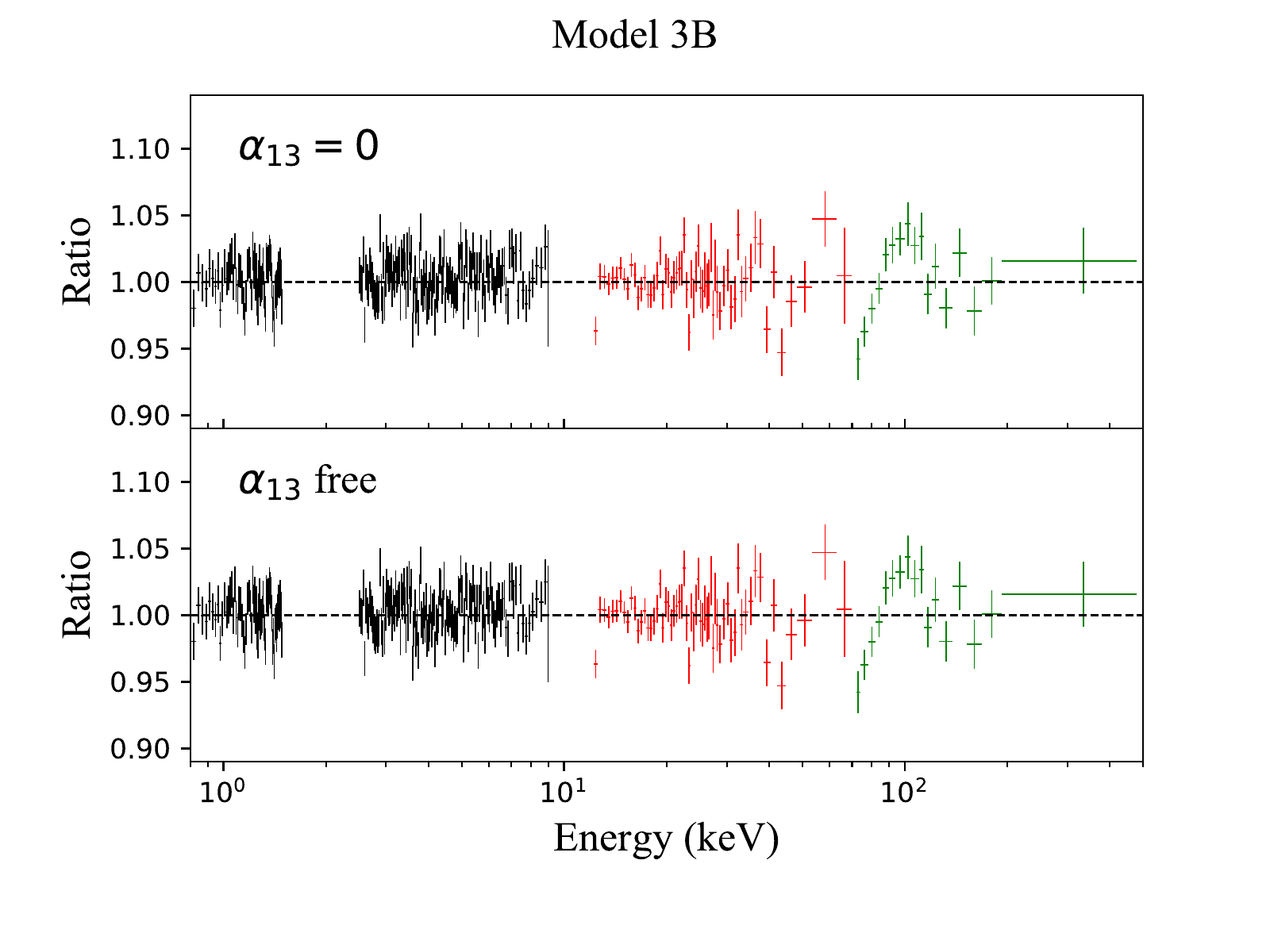} \\
	\end{center}
	\vspace{-0.8cm}
	\caption{Data to best-fit model ratios for Models~1A and 1B (top panels), Models~2A and 2B (central panels), and Models~3A and 3B (bottom panels). For every model, we show the results from the fit with $\alpha_{13} = 0$ and from the fit with $\alpha_{13}$ free.
	\label{f-c2c3c4}}
\end{figure*}

\begin{figure*}[t]
	\begin{center}
		\includegraphics[width=8.5cm,trim={2.5cm 2.5cm 2.5cm 2.5cm},clip]{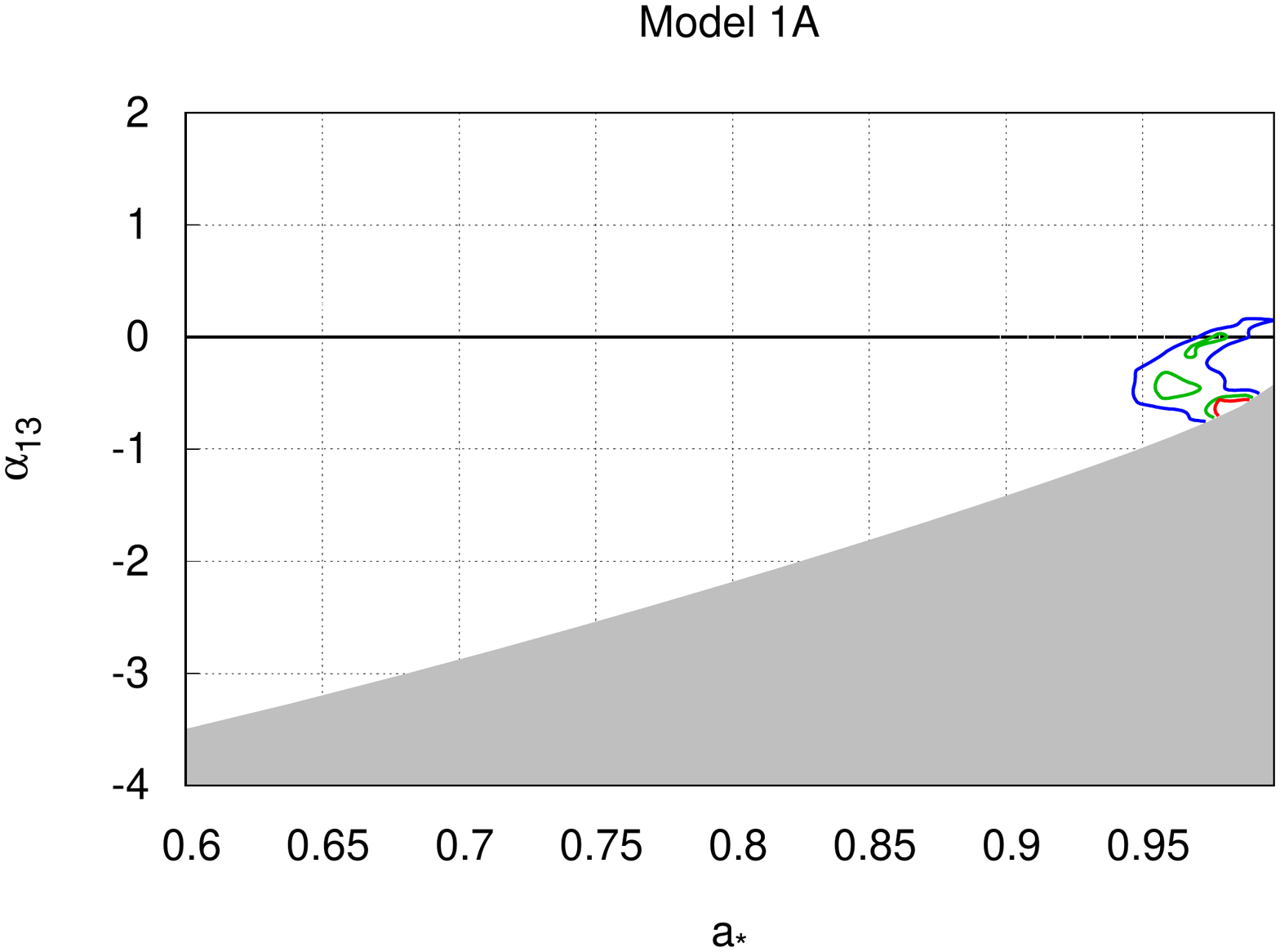}
		\includegraphics[width=8.5cm,trim={2.5cm 2.5cm 2.5cm 2.5cm},clip]{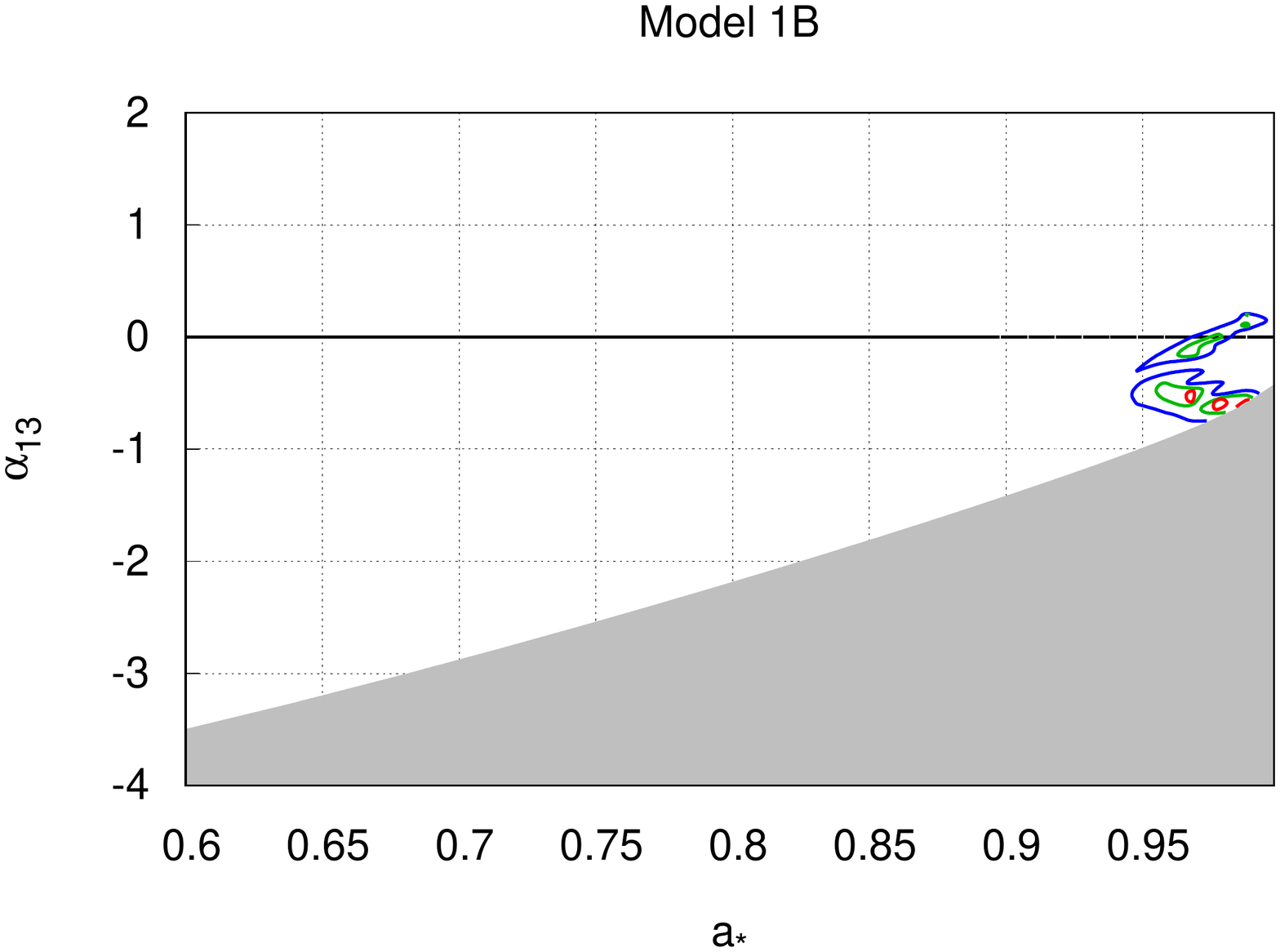} \\
		\includegraphics[width=8.5cm,trim={2.5cm 2.5cm 2.5cm 2.5cm},clip]{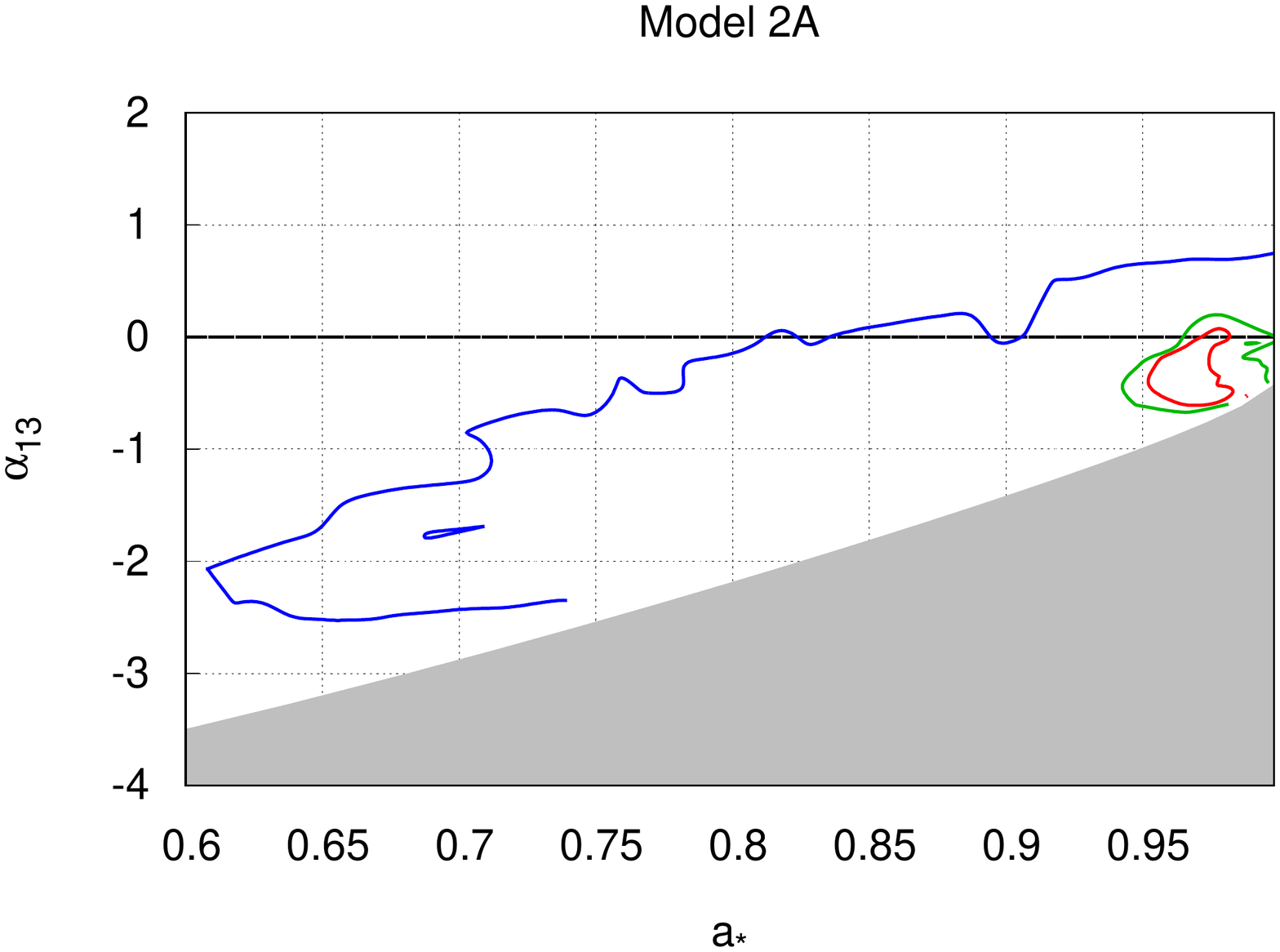}
		\includegraphics[width=8.5cm,trim={2.5cm 2.5cm 2.5cm 2.5cm},clip]{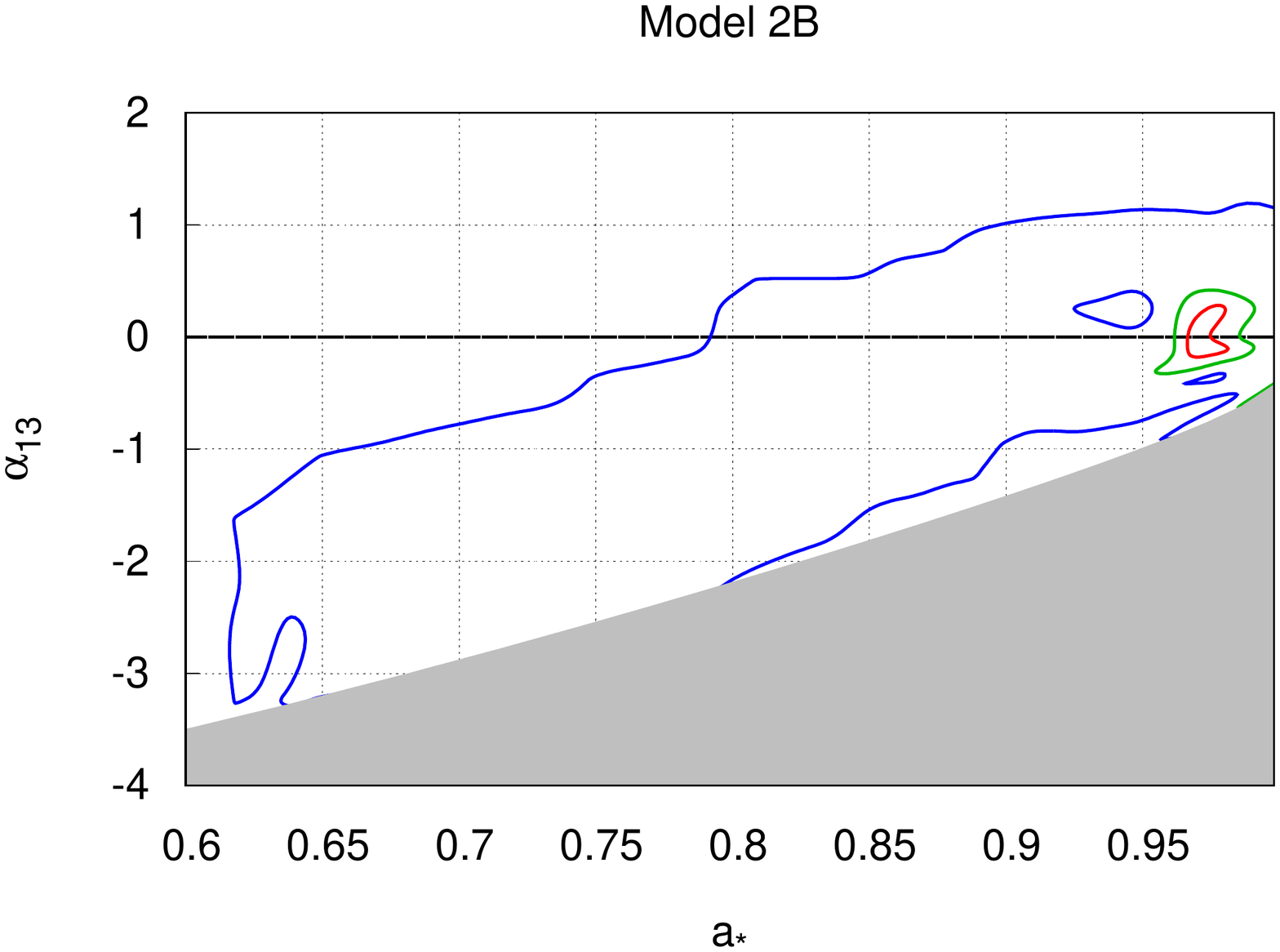} \\
		\includegraphics[width=8.5cm,trim={2.5cm 2.5cm 2.5cm 2.5cm},clip]{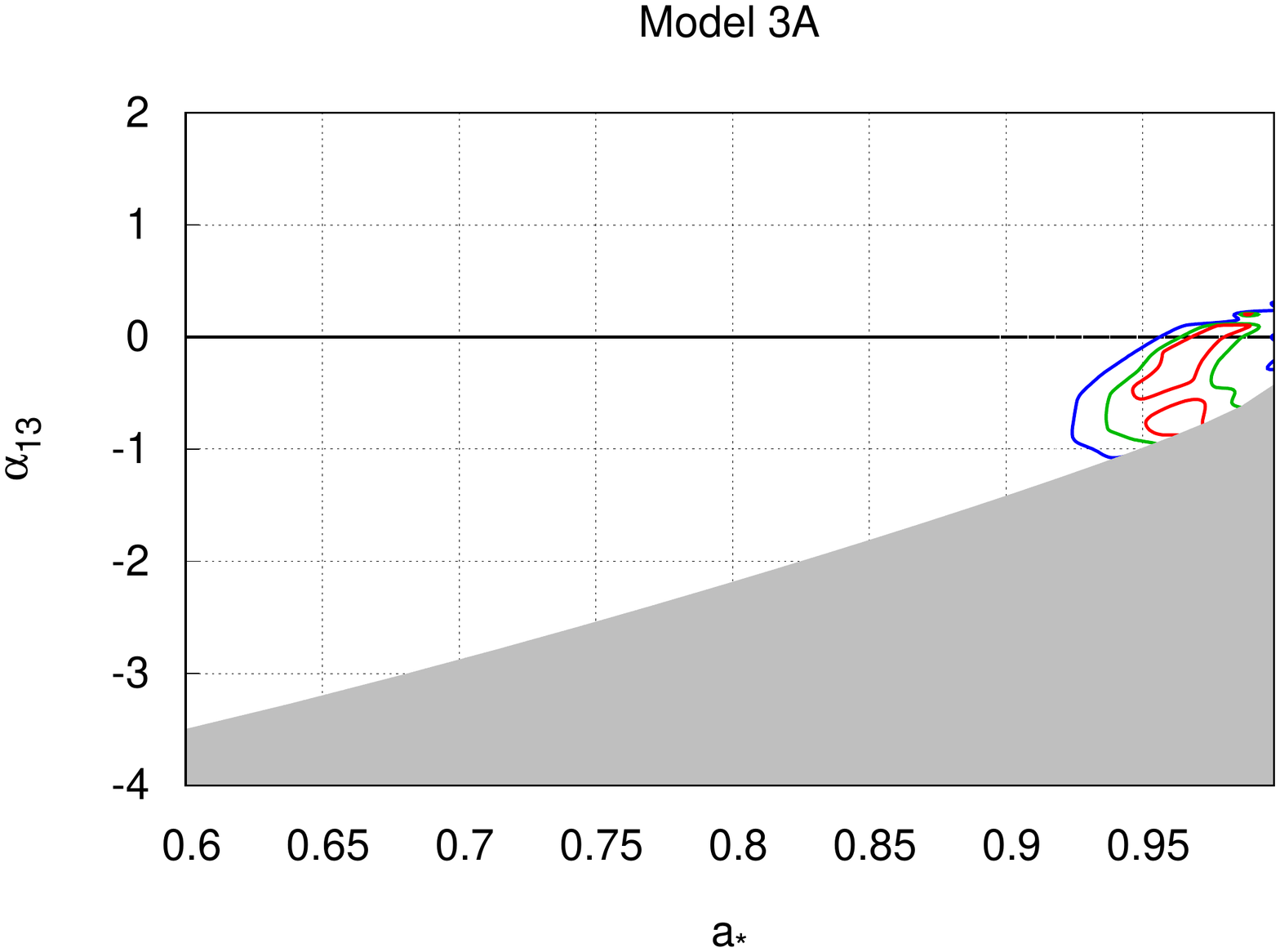}
		\includegraphics[width=8.5cm,trim={2.5cm 2.5cm 2.5cm 2.5cm},clip]{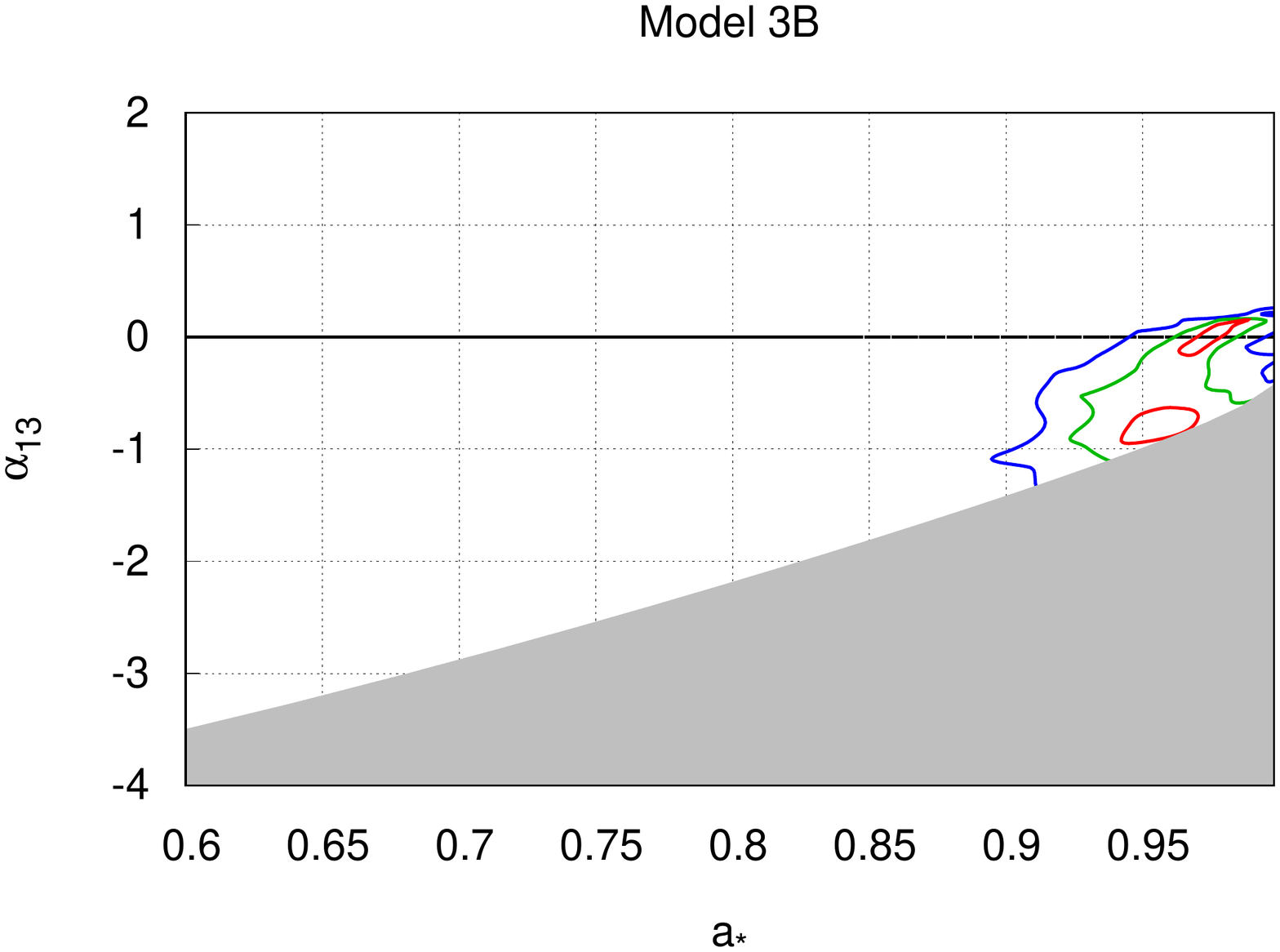} \\
	\end{center}
	\vspace{-0.8cm}
	\caption{Constraints on the spin parameter $a_*$ and the Johannsen deformation parameters $\alpha_{13}$ for Models~1A and 1B (top panels), Models~2A and 2B (central panels), and Models~3A and 3B (bottom panels). The red, green, and blue curves are, respectively, the 68\%, 90\%, and 99\% confidence level boundaries for two relevant parameters. The gray region is ignored in our analysis because the spacetime is not regular there. 
	\label{f-c5}}
\end{figure*}

If $N_{\rm H}$ in {\tt tbabs} is left free, the fit gives us the value $\approx6\times10^{21}$~cm$^{-2}$, which is perfectly consistent with the result in Ref.~\cite{nh}. So we freeze $N_{\rm H}$ to that value for Model~1. 
The best-fit values for Models~1A and 1B are reported in Tab.~\ref{bestfit_1}. The data to best-fit model ratio of these models are shown in the top panels in Fig.~\ref{f-c2c3c4}. The top panels in Fig.~\ref{f-c5} show the constraints on the black hole spin parameter $a_*$ and the Johannsen deformation parameter $\alpha_{13}$ when $\alpha_{13}$ is free in the fit.

Following Refs.~\cite{Miller:2012zj,c2}, we try to improve the fit by replacing {\tt diskbb} with {\tt diskpbb}. In {\tt diskpbb}, the local disk temperature $T(r)$ is proportional to $r^{-p}$, where $p$ is the extra parameter of the model. The spectrum of {\tt diskbb} is recovered for $p = 0.75$. Model~2 is thus

\vspace{0.2cm}

\noindent {\tt tbabs$\times$xstar$\times$(diskpbb+cutoffpl+gauss+relxill\_nk)},

\vspace{0.2cm}

\noindent and we set $p = 0.5$ as in Refs.~\cite{Miller:2012zj,c2}. The idea is that in the hard state the corona can significantly illuminate the disk and thus alter the emissivity profile predicted by the standard thin disk model. As for Model~1, we freeze the column density in {\tt tbabs} to $6 \cdot 10^{21}$~cm$^{-2}$, we assume an emissivity profile described by a broken power-law with $q_{\rm out}$ either frozen to 3 (Model~2A) or free (Model~2B), and for every case we first fit the data imposing the Kerr metric and then we leave $\alpha_{13}$ free in the fit. The best-fit values are shown in Tab.~\ref{bestfit_2}. In the central panels of Fig.~\ref{f-c2c3c4}, we show the data to best-fit model ratios for Model~2A and 2B. In the central panels of Fig.~\ref{f-c5}, we show the constraints on the black hole spin and deformation parameter $\alpha_{13}$. We note that Model~2 improve the fit with respect to Model~1 and $\Delta \chi^2$ is around 80.

Motivated by the fact that both Model~1 and Model~2 require a high value for $q_{\rm in}$ and a value around 3 for $q_{\rm out}$ when the latter is left free, we try to improve our fit replacing {\tt relxill\_nk} with {\tt relxilllp\_nk}, namely replacing a broken power-law emissivity profile with the profile expected in the lamppost geometry. However, our fits turn out to be slightly worse than those from Model~2B ($\chi_{\nu}^2 = 1.075$) and we do not report the results here. We just note that the best-fit values of the model parameters are very similar to those from Model~2B, with the exception of $T_{\rm in}$, which is higher.

Last, we try to improve the fit by modeling better the Comptonized photons of the corona. We replace {\tt cutoffpl+diskpbb} with {\tt simplcutx$\times$diskpbb} and {\tt relxill\_nk} with {\tt relxillCp\_nk}. {\tt simplcutx} is a convolution model that requires as input the seed photon spectrum responsible for the Comptonization and its output includes both the Comptonized photon spectrum and the transmitted seed photon spectrum~\cite{Steiner:2017bhe}. In {\tt simplcutx}, we set the reflection fraction parameter to $1$ and we link the value of the coronal temperature with its counterpart in {\tt relxillCp\_nk}. Our Model~3 is thus

\vspace{0.2cm}

\noindent {\tt tbabs$\times$xstar$\times$(simplcutx$\times$diskpbb+gauss+relxillCp\_nk)}.

\vspace{0.2cm}

\noindent Unlike in Models~1 and 2, here the value of the hydrogen column density $N_{\rm H}$ in {\tt tbabs} has a stronger impact on the final fit. We thus set $N_{\rm H}$ free in the fit. As for Models~1 and 2, we consider the two variants with $q_{\rm out} = 3$ (3A) and free (3B), and in both cases we fit the data first assuming the Kerr metric and then relaxing this hypothesis. The best-fit values of Model~3 are reported in Tab.~\ref{bestfit_3}, where we can see that the fit is better than Model~2 ($\chi_{\nu}^2=1.046$). The data to best-fit model ratios are shown in Fig.~\ref{f-c2c3c4} and the constraints on the black hole spin and the Johannsen deformation parameter are reported in the bottom panels in Fig.~\ref{f-c5}.

The discussion of the fits of epoch~1 is postponed to Section~\ref{s-d-c}.


\section{Analysis of epoch~2 \label{s-ana-s}}

To simplify the analysis, we first fit the \textsl{NuSTAR} and \textsl{Suzaku} data separately, and then we combine the two spectra for a joint fit. When we fit the spectra separately, we set $N_{\rm H}$ in {\tt tbabs} to $6\times10^{21}$~cm$^{-2}$, as in Models~1 and 2 in the previous section. When we fit the \textsl{NuSTAR} and \textsl{Suzaku} data together, $N_{\rm H}$ has a more important influence on the final result and we do not recover $6\times10^{21}$~cm$^{-2}$ when it is allowed to vary, so we leave the parameter free in the fit.
As in the previous section for the analysis of epoch~1, we consider two models for the intensity profile: broken power-law with $q_{\rm out}=3$ and broken power-law with $q_{\rm out}$ free. For every model of the intensity profile, we consider two spacetimes: Kerr metric and Johannsen metric with $\alpha_{13}$ free.

We start with the analysis of the \textsl{NuSTAR} data and we use the XSPEC model

 \vspace{0.2cm}

\noindent {\tt tbabs$\times$xstar$\times$(diskbb+cutoffpl+gauss+relxill\_nk)}.

\vspace{0.2cm}

\begin{figure*}[t]
	\begin{center}
		\includegraphics[scale=0.54]{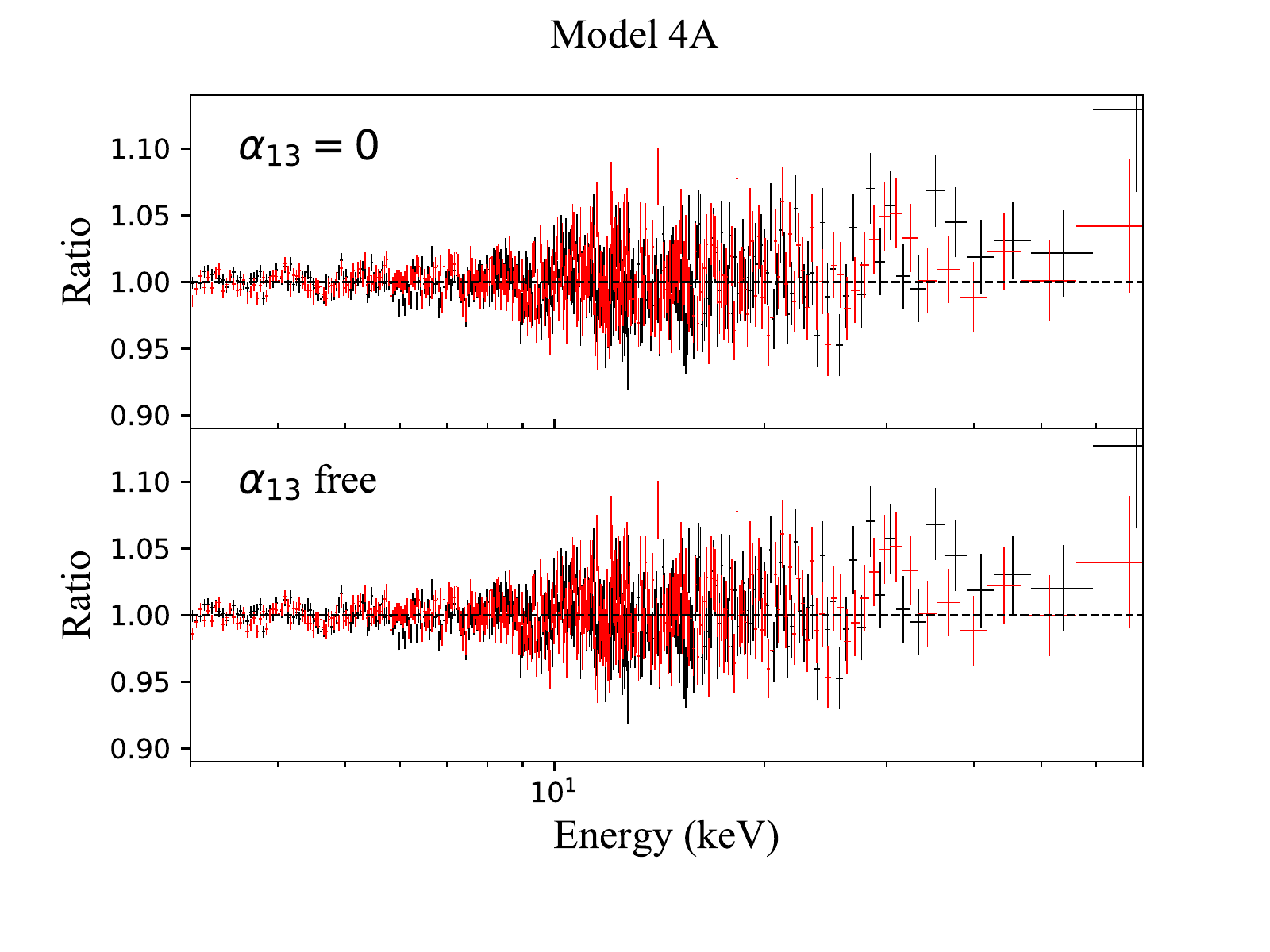}
		\includegraphics[scale=0.54]{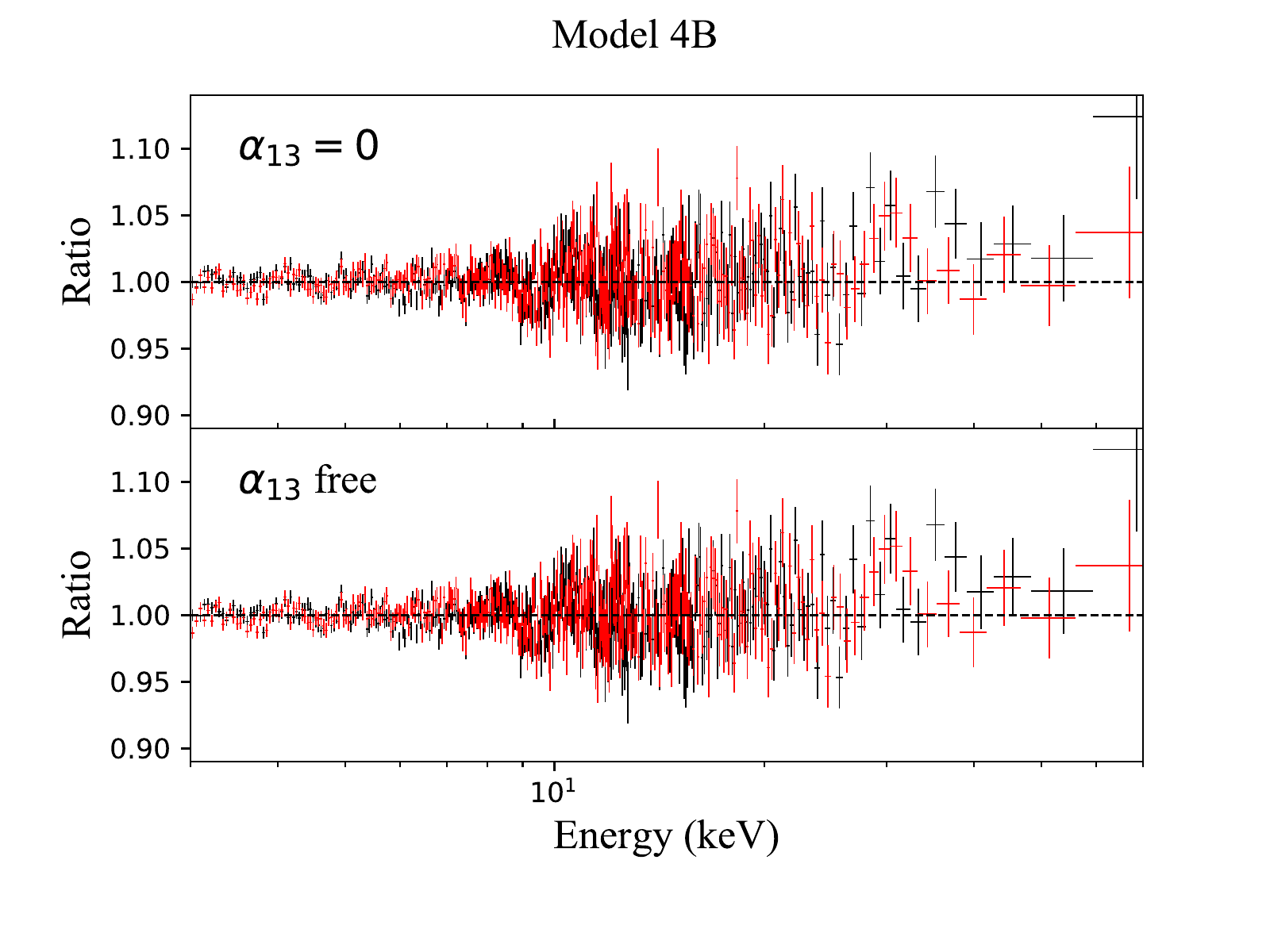} \\
	\end{center}
	\vspace{-1.0cm}
	\caption{Data to best-fit model ratios for Models~4A and 4B (\textsl{NuSTAR} data of epoch~2). For every model, we show the results from the fit with $\alpha_{13} = 0$ and from the fit with $\alpha_{13}$ free.
	\label{f-c6}}
\end{figure*}

\begin{figure*}[t]
	\begin{center}
		\includegraphics[width=8.5cm,trim={2.5cm 2.5cm 2.5cm 2.5cm},clip]{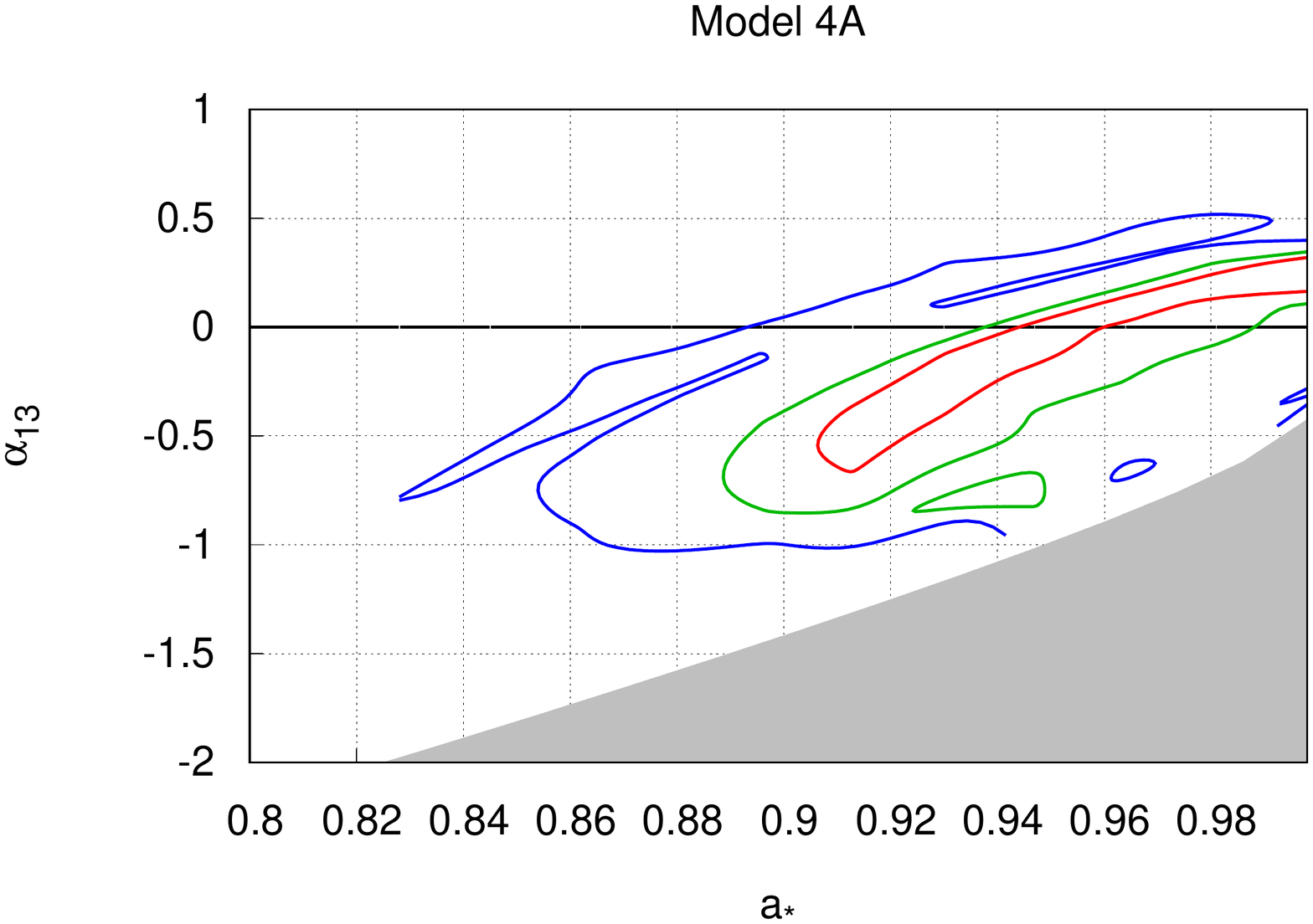}
		\includegraphics[width=8.5cm,trim={2.5cm 2.5cm 2.5cm 2.5cm},clip]{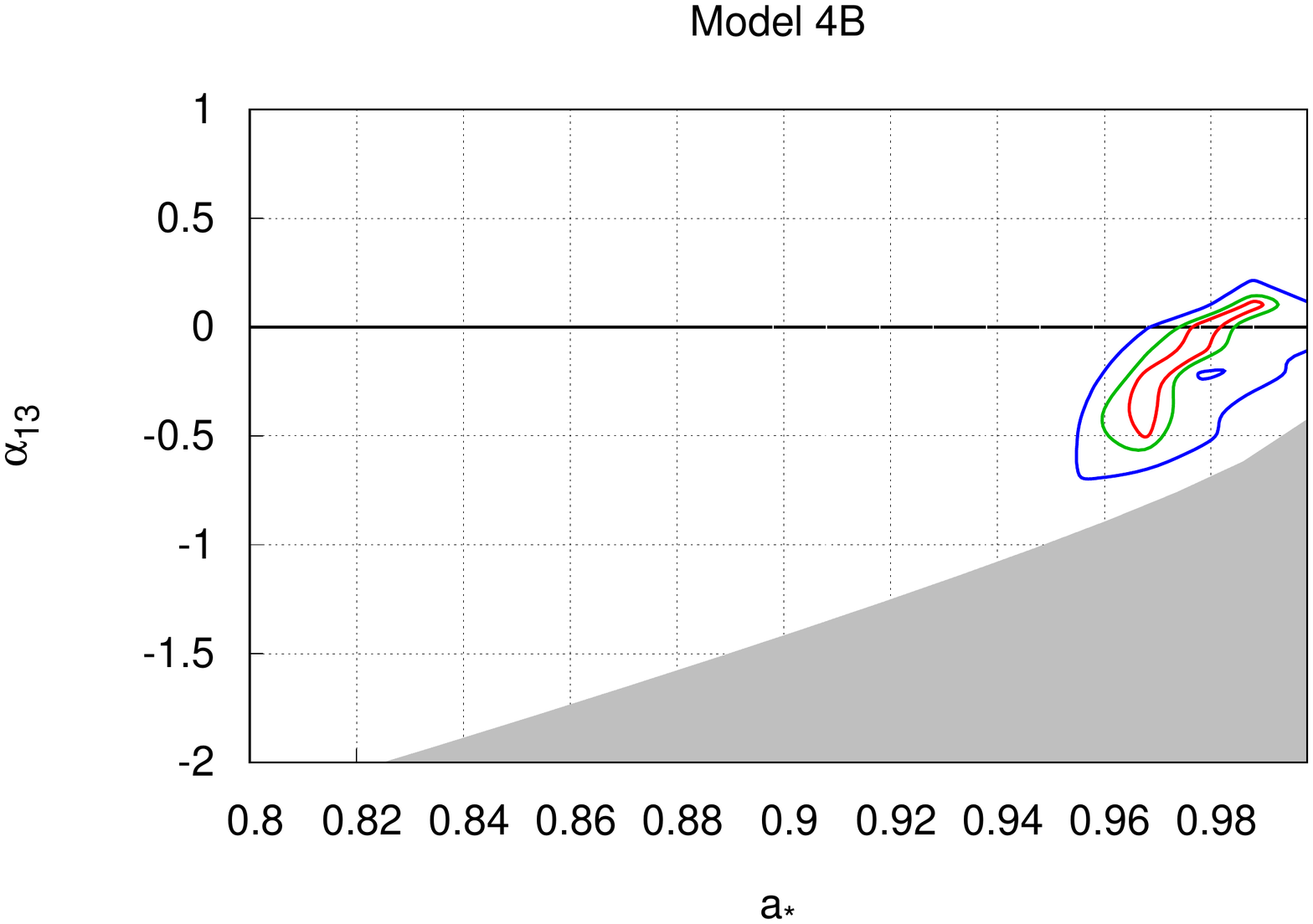} \\
	\end{center}
	\vspace{-0.8cm}
	\caption{Constraints on the spin parameter $a_*$ and the Johannsen deformation parameters $\alpha_{13}$ for Models~4A and 4B (\textsl{NuSTAR} data of epoch~2). The red, green, and blue curves are, respectively, the 68\%, 90\%, and 99\% confidence level boundaries for two relevant parameters. The gray region is ignored in our analysis because the spacetime is not regular there. 
	\label{f-c7}}
\end{figure*}

\noindent This is the same model as Model~1 for epoch~1, but for convenience we call it Model~4 as it fits the \textsl{NuSTAR} data of epoch~2. Following the same strategy employed in the previous section, we describe the emissivity profile with a broken power-law and the outer emissivity profile $q_{\rm out}$ is either frozen to 3 (Model~4A) or left free in the fit (Model~4B). For every model, we first fit the data assuming the Kerr metric ($\alpha_{13} = 0$) and then without such an assumption ($\alpha_{13}$ free). The best-fit values are reported in Tab.~\ref{bestfit_4}. The data to best-fit model ratios are shown in Fig.~\ref{f-c6}. The constraints on the black hole spin parameter $a_*$ and the Johannsen deformation parameter $\alpha_{13}$ are in Fig.~\ref{f-c7}.

Since in this observation the source is in the soft state, the accretion disk is not cold, which is instead an assumption in the reflection model used. As done in Ref.~\cite{w16} and then repeated in Ref.~\cite{Liu:2019vqh}, we try to take the higher disk temperature into account by adding a Gaussian convolution model to the reflection spectrum at the emission point and before the relativistic convolution model. In XSPEC language, we replace {\tt relxill\_nk} with

 \vspace{0.2cm}

\noindent {\tt relconv\_nk$\times$gsmooth$\times$xillver}.

\vspace{0.2cm}

\noindent However, we do not have an appreciable improvement of the fit and, on the contrary, the fit would require a negative value for $q_{\rm out}$. So we do not explore further such a possibility.

We move to the analysis of the \textsl{Suzaku} data and we use the same XSPEC model:

 \vspace{0.2cm}

\noindent {\tt tbabs$\times$xstar$\times$(diskbb+cutoffpl+gauss+relxill\_nk)}.

\vspace{0.2cm}

\begin{figure*}[t]
	\begin{center}
		\includegraphics[scale=0.54]{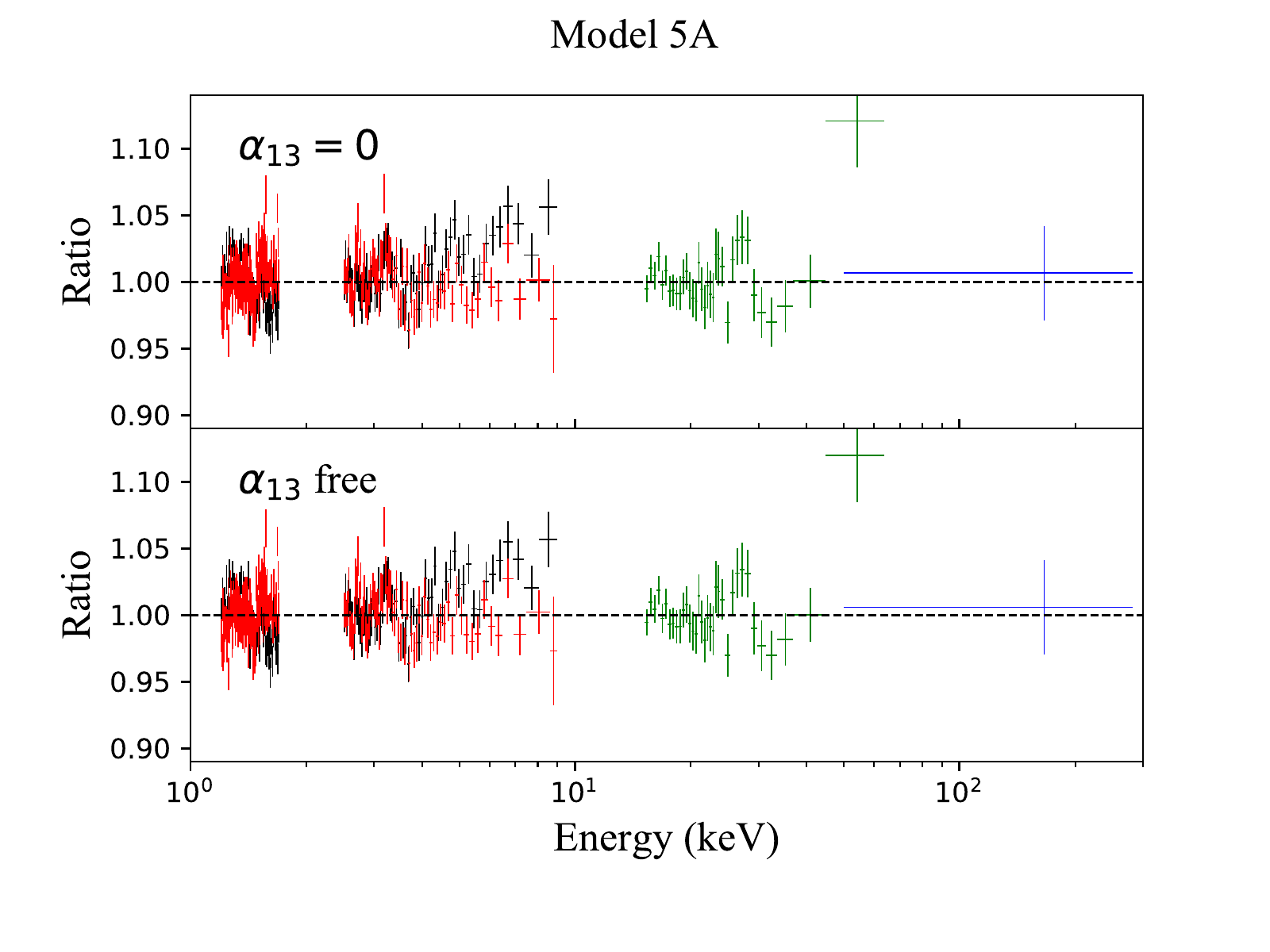}
		\includegraphics[scale=0.54]{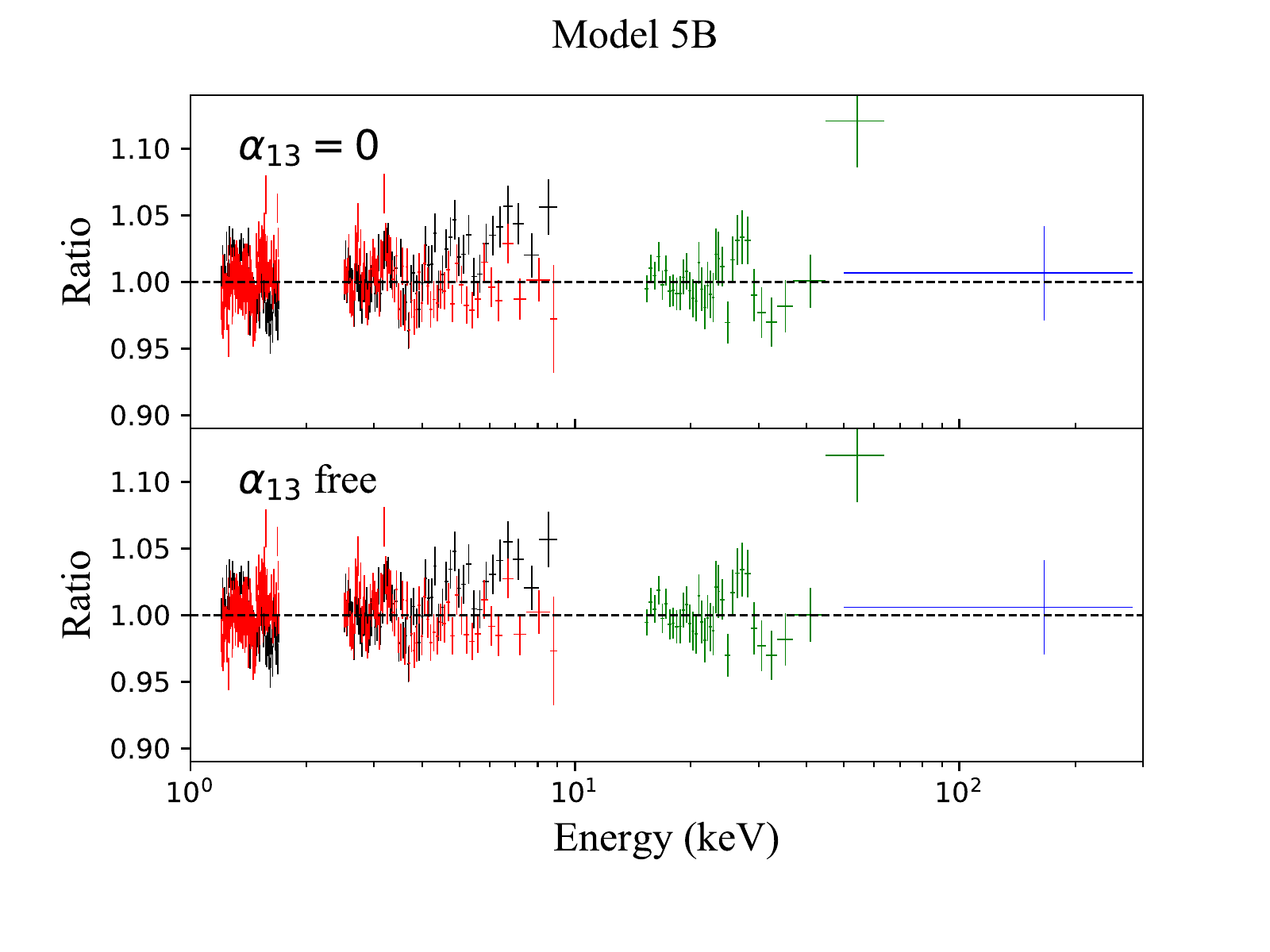} \\
	\end{center}
	\vspace{-1.0cm}
	\caption{Data to best-fit model ratios for Models~5A and 5B (\textsl{Suzaku} data of epoch~2). For every model, we show the results from the fit with $\alpha_{13} = 0$ and from the fit with $\alpha_{13}$ free.
	\label{f-c8}}
\end{figure*}

\begin{figure*}[t]
	\begin{center}
		\includegraphics[width=8.5cm,trim={2.5cm 2.5cm 2.5cm 2.5cm},clip]{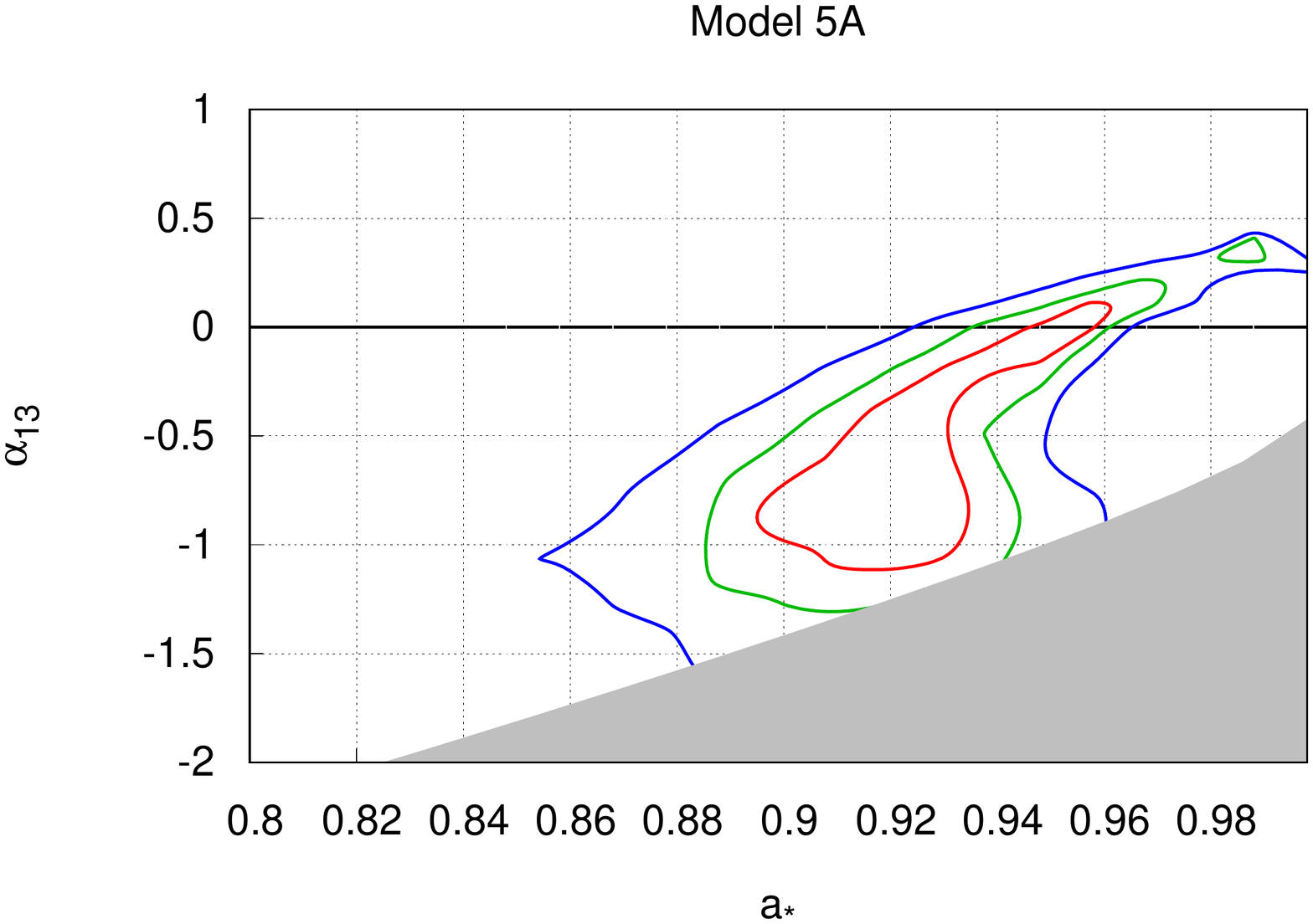}
		\includegraphics[width=8.5cm,trim={2.5cm 2.5cm 2.5cm 2.5cm},clip]{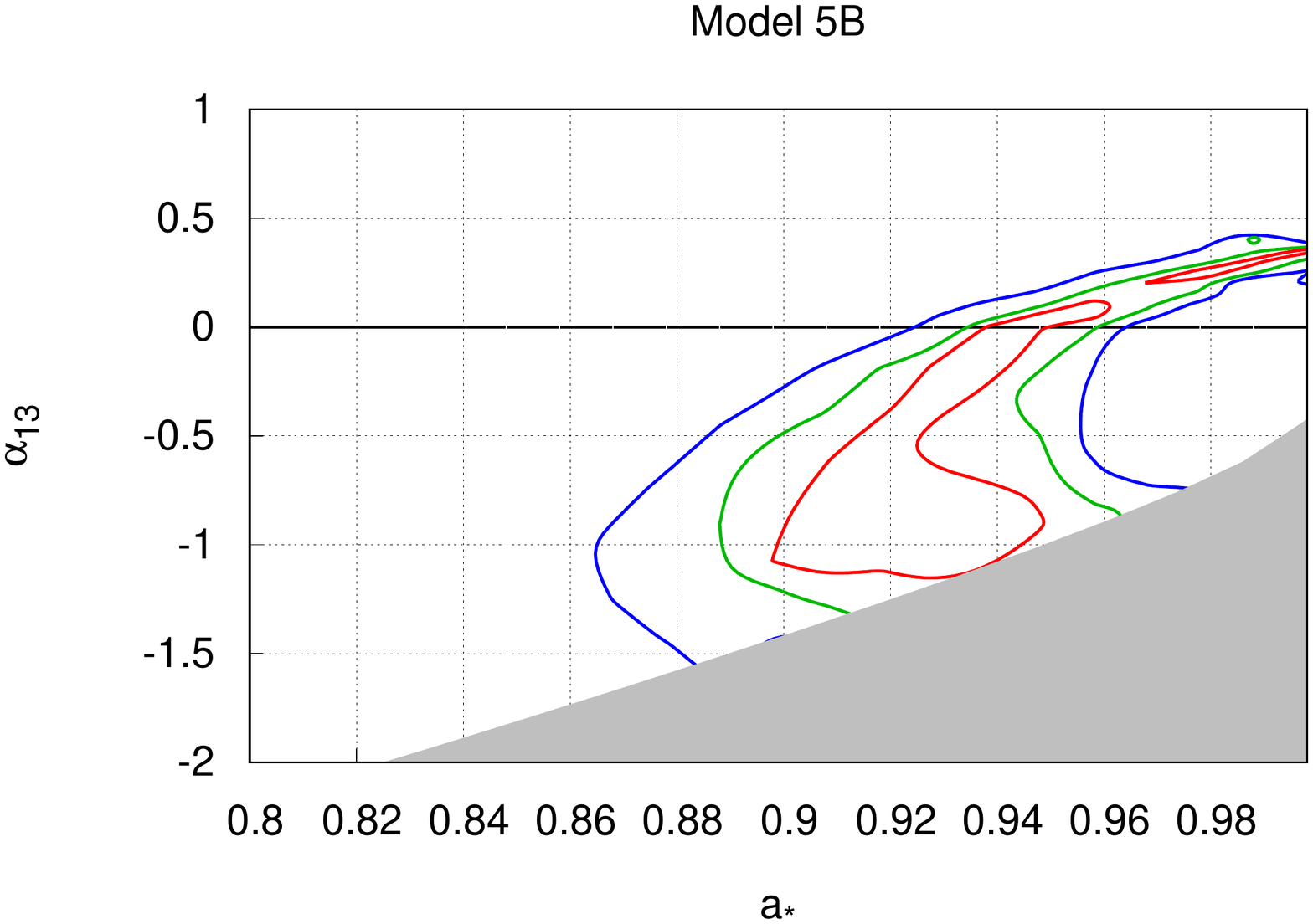} \\
	\end{center}
	\vspace{-0.8cm}
	\caption{Constraints on the spin parameter $a_*$ and the Johannsen deformation parameters $\alpha_{13}$ for Models~5A and 5B (\textsl{Suzaku} data of epoch~2). The red, green, and blue curves are, respectively, the 68\%, 90\%, and 99\% confidence level boundaries for two relevant parameters. The gray region is ignored in our analysis because the spacetime is not regular there. 
	\label{f-c9}}
\end{figure*}

\noindent Since we fit now the \textsl{Suzaku} data, we call this model Model~5, with Model~5A with $q_{\rm out}=3$ and Model~5B with $q_{\rm out}$ free. The best-fit values are reported in Tab.~\ref{bestfit_5}. The data to best-fit model ratios are in Fig.~\ref{f-c8}. The constraints on the black hole spin parameter $a_*$ and the Johannsen deformation parameter $\alpha_{13}$ are shown in Fig.~\ref{f-c9}. We note that there is some disagreement between the XIS0 and XIS1 data. It may be related to some residual pile-up.  
As in the \textsl{NuSTAR} data, we try to improve the fit by replacing {\tt relxill\_nk} with {\tt relconv\_nk$\times$gsmooth$\times$xillver} to take into account the fact that the source is in the soft state and its accretion disk is not cold. However, as in the case of the \textsl{NuSTAR} spectrum, the fit does not improve.

\begin{figure*}[t]
	\begin{center}
		\includegraphics[scale=0.54]{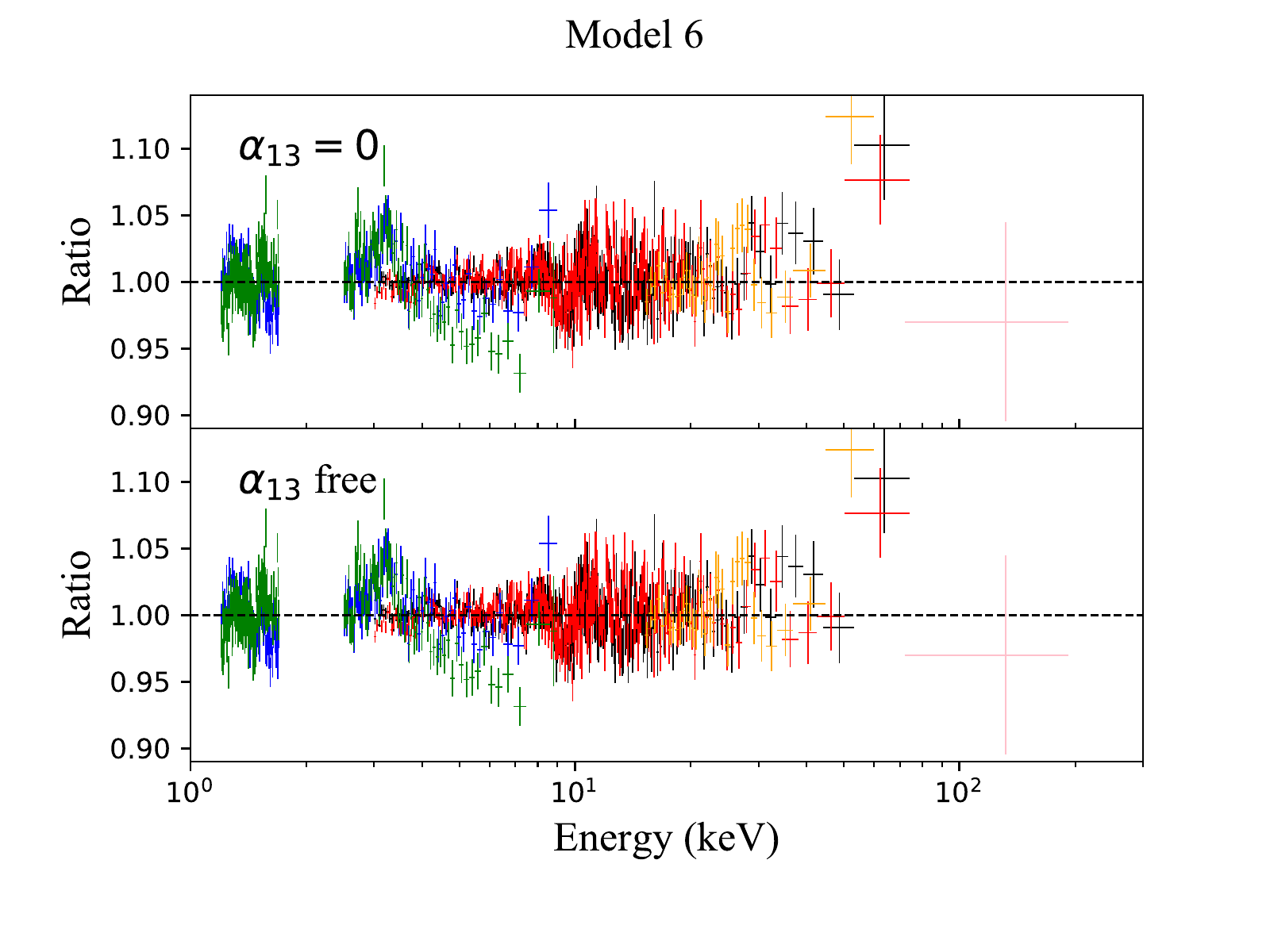}
	\end{center}
	\vspace{-1.0cm}
	\caption{Data to best-fit model ratios for Model~6 (\textsl{NuSTAR}+\textsl{Suzaku} data of epoch~2). The emissivity profile is modeled with broken power-law with $q_{\rm out}=3$. \label{f-c10}}
\end{figure*}

\begin{figure*}[t]
	\begin{center}
		\includegraphics[width=8.5cm,trim={2.5cm 2.5cm 2.5cm 2.5cm},clip]{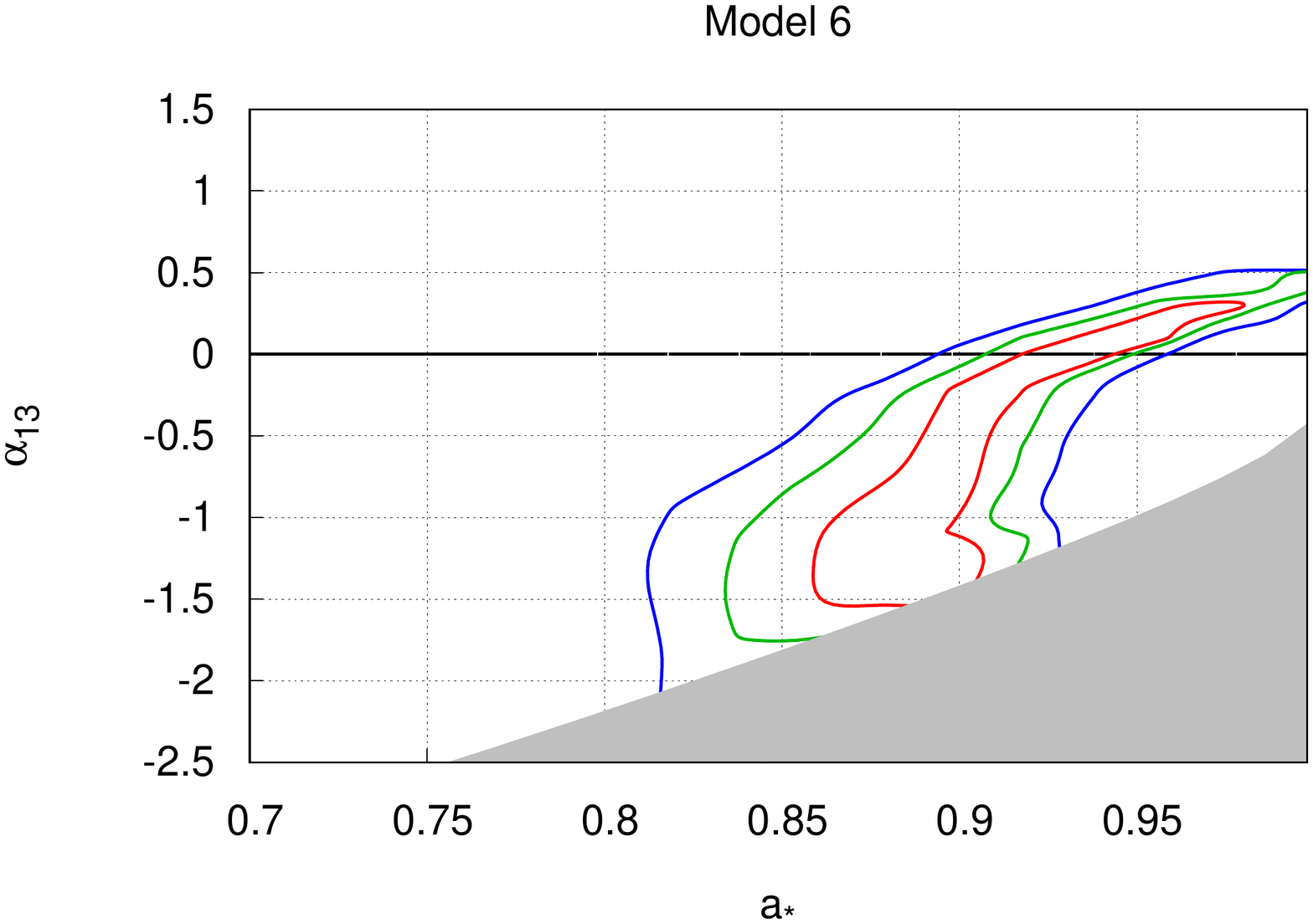}
	\end{center}
	\vspace{-0.8cm}
	\caption{Constraints on the spin parameter $a_*$ and the Johannsen deformation parameters $\alpha_{13}$ for Model~6 (\textsl{NuSTAR}+\textsl{Suzaku} data of epoch~2). The emissivity profile is modeled with a broken power-law with outer emissivity index frozen to 3. The red, green, and blue curves are, respectively, the $68\%$, $90\%$, and $99\%$ confidence level boundaries for two relevant parameters. \label{f-c11}}
\end{figure*}

Last, we fit the \textsl{NuSTAR} and \textsl{Suzaku} spectra together. $N_{\rm H}$ in {\tt tbabs} is left free. Since we cannot constrain $q_{\rm out}$ when it is allowed to vary, we freeze it to 3. This is our Model~6, and now we do not have Models~6A and 6B. The results of our fits are shown in Tab.~\ref{bestfit_6}, Fig.~\ref{f-c10} and Fig.~\ref{f-c11}.


\begin{table*}[]
	\centering
	\vspace{0.5cm}
	\begin{tabular}{lcc|cc}
		\hline\hline
		Model            & \multicolumn{2}{c}{1A} & \multicolumn{2}{c}{1B} \\
		& $\alpha_{13}=0$ & $\alpha_{13}$ free & $\alpha_{13}=0$ & $\alpha_{13}$ free \\ 
		\hline
		{\tt tbabs}   \\
		$ N_{\rm H}/10^{21}$~cm$^{-2}$ &  $6.0^*$         &  $6.0^*$  &  $6.0^*$         &  $6.0^*$   \\ \hline
		{\tt xstar}  \\ 
		$ N_{\rm H}/10^{21}$~cm$^{-2}$ & $12.5_{-6}^{+1.2} $ & $10_{-4}^{+4} $ & $9_{-5}^{+11} $ & $9_{-5}^{+11} $  \\ 
		$\log\xi$ &   $3.48_{-0.11}^{+0.4}$  &   $3.44_{-0.06}^{+0.23}$   &   $3.44_{-0.07}^{+0.6}$ &   $3.39_{-0.18}^{+0.29}$  \\ 
		$ z $& $-0.017\pm0.006 $ & $-0.016_{-0.006}^{+0.007} $  & $-0.016_{-0.008}^{+0.007} $ & $-0.017\pm0.006 $ \\ \hline
		{\tt relxill\_nk}  \\ 
		$ q_{\rm in} $&   $>7.4$   &   $8.92_{-1.2}^{+0.22}$ &   $>7.8$  &   $>8.7$      \\ 
		$ q_{\rm out} $&   $3^*$ &   $3^*$ &   $3.16_{-0.3}^{+0.24}$ &   $3.08_{-0.07}^{+0.03}$ \\ 
		$ R_{\rm br} $ [$r_{\rm g}$]&    $2.73_{-0.20}^{+0.28}$ &   $1.89_{-0.08}^{+0.02}$  &   $2.50_{-0.29}^{+0.5}$ &   $1.78_{-0.04}^{+0.02}$ \\ 
		$ a_* $&     $0.976\pm0.005$ &   $0.988_{-0.009}^{+0.0012}$  &$0.976_{-0.003}^{+0.005}$&$0.988_{-0.009}^{+0.0012}$ \\ 
		$ i $ [deg]&     $40.0_{-1.4}^{+1.5}$ & $40.0_{-1.1}^{+0.7}$& $40.1_{-1.0}^{+2.0}$ & $40.3_{-1.3}^{+0.6}$ \\ 
		$\log\xi $&     $3.13_{-0.04}^{+0.03}$ & $3.136_{-0.009}^{+0.023}$& $3.12_{-0.03}^{+0.05}$ & $3.150_{-0.011}^{+0.015}$ \\ 
		$ A_{\rm Fe} $&     $>9.64$ & $>9.79$ & $>7.76$ & $>9.81$ \\ 
		$ \alpha_{13} $&       $0^*$    &  $<-0.6$ &  $0^*$ &  $<-0.5$  \\ 
		Norm&   $0.0134_{-0.0011}^{+0.0008}$ &  $0.0132_{-0.0003}^{+0.0006}$ &  $0.0136_{-0.0010}^{+0.0009}$ &  $0.0133_{-0.0003}^{+0.0006}$ \\ \hline
		{\tt cutoffpl} \\ 
		$ \Gamma $& $1.557_{-0.014}^{+0.011} $ &   $1.555_{-0.008}^{+0.006}$  &   $1.556_{-0.016}^{+0.016}$ &   $1.548_{-0.007}^{+0.008}$    \\ 
		$ E_{\rm cut} $ [keV] &  $286_{-12}^{+42}$  & $284_{-3}^{+16}$& $287_{-20}^{+37}$ & $278_{-3}^{+4}$  \\ 
		Norm&      $0.98\pm0.04$ &  $0.981_{-0.022}^{+0.013}$ &  $0.98_{-0.05}^{+0.03}$ &  $0.955_{-0.019}^{+0.04}$ \\ \hline
		{\tt diskbb} \\ 
		$ T_{\rm in} $ [keV] &    $0.58\pm0.03$ & $0.568_{-0.003}^{+0.021}$& $0.58\pm0.03$ & $0.554_{-0.003}^{+0.011}$   \\ 
		Norm&     $401_{-65}^{+95}$ & $432_{-9}^{+76}$& $419_{-123}^{+103}$ & $482_{-9}^{+60}$       \\ \hline
		{\tt gauss} \\ 
		$E_{\rm line}$ [keV] &       $6.40^*$  & $6.40^*$ & $6.40^*$  & $6.40^*$  \\ 
		$ \sigma $ [keV] &     $0.01^*$  & $0.01^*$  & $0.01^*$  & $0.01^*$   \\ 
		Norm/$10^{-3}$& $0.95\pm0.27$ & $0.94_{-0.23}^{+0.25}$& $0.98_{-0.26}^{+0.27}$ & $0.99_{-0.23}^{+0.24}$ \\ \hline
		Cross-normalization \\
		$C_{\rm PIN}$ & $1.225_{-0.028}^{+0.027}$ & $1.227_{-0.006}^{+0.004}$ & $1.219_{-0.025}^{+0.029}$ & $1.223_{-0.008}^{+0.004}$ \\
		$C_{\rm GSO}$ & $1.43\pm0.04$ & $1.439_{-0.008}^{+0.015}$ & $1.43_{-0.03}^{+0.04}$ & $1.439_{-0.008}^{+0.015}$ \\ \hline
		$\chi^2/\nu $ & $2373.82/2139$ & $2371.13/2138$ & $2372.82/2138$ & $2370.01/2137$ \\ 
		& $=1.10978$ & $=1.10904$ & $=1.10983$ & $=1.10903$ \\
		\hline\hline
	\end{tabular}
	\vspace{0.2cm}
	\caption{Best-fit values from Models~1A and 1B. The reported uncertainties correspond to the 90\% confidence level for one relevant parameter. $^*$ means that the parameter is frozen in the fit. The ionization parameter $\xi$ is in units erg~cm~s$^{-1}$.  
	\label{bestfit_1}}
\end{table*}

\begin{table*}[]
	\centering
	\vspace{0.5cm}
	\begin{tabular}{lcc|cc}
		\hline\hline
		Model            & \multicolumn{2}{c}{2A} & \multicolumn{2}{c}{2B} \\
		& $\alpha_{13}=0$ & $\alpha_{13}$ free & $\alpha_{13}=0$ & $\alpha_{13}$ free \\ 
		\hline
		{\tt tbabs} \\
		$ N_{\rm H}/10^{21}$~cm$^{-2}$ &  $6.0^*$         &  $6.0^*$  &  $6.0^* $        &  $6.0^* $  \\ \hline
		{\tt xstar} \\ 
		$ N_{\rm H}/10^{21}$~cm$^{-2}$ & $10_{-6}^{+9} $ & $10_{-6}^{+9} $ & $12_{-6}^{+9} $ & $12_{-6}^{+9} $  \\ 
		$\log\xi $&   $3.40_{-0.24}^{+0.4}$  &   $3.39_{-0.24}^{+0.4}$   &   $3.39_{-0.18}^{+0.29}$ &   $3.39_{-0.18}^{+0.29}$  \\ 
		$ z $& $-0.017\pm0.006 $ & $-0.017\pm0.006 $  & $-0.017\pm0.006 $ & $-0.017_{-0.005}^{+0.006} $ \\ \hline
		{\tt relxill\_nk} \\ 
		$ q_{\rm in} $&   $>7.3$   &   $>8.5$ &   $>7.3$  &   $>7.0$      \\ 
		$ q_{\rm out} $&   $3^*$ &   $3^*$ &   $2.86_{-0.18}^{+0.15}$ &   $2.86_{-0.3}^{+0.15}$ \\ 
		$ R_{\rm br} $ [$r_{\rm g}$] &    $2.17_{-0.3}^{+0.13}$ &   $2.21_{-0.25}^{+0.13}$  &   $2.21_{-0.21}^{+0.23}$ &   $2.21_{-0.7}^{+0.29}$ \\ 
		$ a_* $&     $0.979_{-0.008}^{+0.006}$ &   $>0.960$  &$0.977_{-0.008}^{+0.007}$&$0.977_{-0.029}^{+0.018}$ \\ 
		$ i $ [deg]&     $42.1_{-1.5}^{+1.2}$ & $42.2_{-1.6}^{+1.2}$& $41.8_{-1.6}^{+1.3}$ & $41.8_{-1.6}^{+1.3}$ \\ 
		$\log\xi $& $3.000_{-0.023}^{+0.011}$ & $3.000_{-0.021}^{+0.010}$ &  $3.000_{-0.024}^{+0.012}$ & $3.000_{-0.024}^{+0.012}$ \\ 
		$ A_{\rm Fe} $&     $5.3_{-0.5}^{+1.2}$ & $5.1_{-0.3}^{+1.2}$ & $5.02_{-0.21}^{+1.2}$ & $5.02_{-0.22}^{+1.12}$ \\ 
		$ \alpha_{13} $&       $0^*$    &  $-0.05_{-0.6}^{+0.18}$ &  $0^*$ &  $0.0_{-0.2}^{+0.3}$  \\ 
		Norm&   $0.0100_{-0.0009}^{+0.0008}$ &  $0.0098_{-0.0009}^{+0.0012}$ &  $0.0094_{-0.0008}^{+0.0011}$ &  $0.0094_{-0.0008}^{+0.0011}$ \\ \hline
		{\tt cutoffpl} \\ 
		$ \Gamma $& $1.542_{-0.009}^{+0.009}$ &   $1.541_{-0.009}^{+0.009}$  &   $1.546_{-0.015}^{+0.011}$ &   $1.546_{-0.013}^{+0.008}$    \\ 
		$ E_{\rm cut}$ [keV]  &  $260_{-7}^{+16}$  & $259_{-6}^{+17}$& $262_{-7}^{+14}$ & $262_{-10}^{+14}$  \\ 
		Norm&      $1.059_{-0.022}^{+0.03}$ &  $1.058_{-0.023}^{+0.022}$ &  $1.068_{-0.029}^{+0.012}$ &  $1.068_{-0.03}^{+0.013}$ \\ \hline
		{\tt diskpbb} \\ 
		$ T_{\rm in} $ [keV] &    $1.29_{-0.09}^{+0.06}$ & $1.29_{-0.03}^{+0.06}$& $1.32_{-0.03}^{+0.16}$ & $1.32_{-0.16}^{+0.02}$   \\ 
		$ p $&       $0.5^*$ &       $0.5^*$ & $0.5^*$ & $0.5^*$    \\ 
		Norm&     $7.7_{-1.5}^{+1.2}$ & $7.7_{-1.4}^{+1.3}$& $6.8_{-1.0}^{+0.6}$ & $6.8_{-0.5}^{+0.7}$       \\ \hline
		{\tt gauss} \\ 
		$E_{\rm line}$ [keV] &       $6.40^*$  & $6.40^*$ & $6.40^*$  & $6.40^*$  \\ 
		$ \sigma $ [keV] &     $0.01^*$  & $0.01^*$  & $0.01^*$  & $0.01^*$   \\ 
		Norm/$10^{-3}$& $1.10_{-0.26}^{+0.25}$ & $1.09_{-0.25}^{+0.24}$& $1.01_{-0.28}^{+0.26}$ & $1.01_{-0.27}^{+0.26}$ \\ \hline
		Cross-normalization \\
		$C_{\rm PIN}$ & $1.195_{-0.010}^{+0.017}$ & $1.195_{-0.009}^{+0.016}$ & $1.207_{-0.013}^{+0.015}$ & $1.207_{-0.009}^{+0.016}$ \\
		$C_{\rm GSO}$ & $1.346_{-0.022}^{+0.026}$ & $1.345_{-0.021}^{+0.026}$ & $1.357_{-0.023}^{+0.024}$ & $1.357_{-0.022}^{+0.024}$ \\ \hline
		$\chi^2/\nu $ & $2294.35/2139$ & $2394.10/2138$  & $2293.08/2138$ & $2293.08/2137$ \\ 
		& $=1.07263$ & $=1.07301$ & $=1.07253$ & $=1.07304$ \\
		\hline\hline
	\end{tabular}
	\vspace{0.2cm}
	\caption{Best-fit values from Models~2A and 2B. The reported uncertainties correspond to the 90\% confidence level for one relevant parameter. $^*$ means that the parameter is frozen in the fit. The ionization parameter $\xi$ is in units erg~cm~s$^{-1}$. 
	 \label{bestfit_2}}
\end{table*}

\begin{table*}[]
	\centering
	\vspace{0.5cm}
	\begin{tabular}{lcc|cc}
		\hline\hline
		Model            & \multicolumn{2}{c}{3A} & \multicolumn{2}{c}{3B} \\
		& $\alpha_{13}=0$ & $\alpha_{13}$ free & $\alpha_{13}=0$ & $\alpha_{13}$ free \\ 
		\hline
		{\tt tbabs} \\
		$ N_{\rm H}/10^{21}$~cm$^{-2}$ &  $5.51_{-0.14}^{+0.13}$   &  $5.51_{-0.14}^{+0.20}$  & $5.51_{-0.15}^{+0.13}$  & $5.51_{-0.21}^{+0.15}$    \\ \hline
		{\tt xstar} \\ 
		$ N_{\rm H}/10^{21}$~cm$^{-2}$ & $6.7\pm0.5 $ & $6.9\pm0.6 $ & $6.8_{-0.4}^{+0.5} $ & $7.1\pm0.5$  \\ 
		$\log\xi $&   $3.4\pm0.3$  &   $3.39_{-0.24}^{+0.3}$   &   $3.4\pm0.3$ &   $3.4_{-0.3}^{+0.4}$  \\ 
		$ z $& $-0.017_{-0.005}^{+0.006} $ & $-0.017_{-0.006}^{+0.004}$  & $0.017\pm0.006$ & $0.017\pm0.005$ \\ \hline
		{\tt relxillCp\_nk}  \\ 
		$ q_{\rm in} $&   $>8.1$   &   $9.2_{-3}^{+0.4}$ &   $>8.3$  &   $>6.1$      \\ 
		$ q_{\rm out} $&   $3^*$ &   $3^*$ &   $3.02_{-0.3}^{+0.24}$ &   $2.94_{-0.23}^{+0.25}$ \\ 
		$ R_{\rm br} $ [$r_{\rm g}$] &    $2.51_{-0.14}^{+0.18}$ &   $2.4_{-0.4}^{+0.4}$  &   $2.50\pm0.25$ &   $2.09_{-0.15}^{+0.9}$ \\ 
		$ a_* $&     $0.972_{-0.009}^{+0.007}$ &   $0.972_{-0.009}^{+0.003}$  &$0.972_{-0.012}^{+0.010}$&$0.972_{-0.024}^{+0.017}$ \\ 
		$ i $ [deg]&     $34.5_{-1.6}^{+2.4}$ & $35.6_{-4}^{+2.4}$& $35_{-4}^{+4}$ & $35_{-8}^{+3}$ \\ 
		log$ \xi $&     $3.30_{-0.09}^{+0.05}$ & $3.29_{-0.08}^{+0.06}$& $3.30_{-0.09}^{+0.05}$ & $3.28_{-0.08}^{+0.11}$ \\ 
		$ A_{\rm Fe} $&     $7.5_{-1.1}^{+1.5}$ & $7.1_{-0.8}^{+2}$ & $7.6_{-1.1}^{+1.3}$ & $7.7_{-1.7}^{+1.7}$ \\ 
		$ \alpha_{13} $&       $0^*$    &  $-0.2_{-0.7}^{+0.3}$ &  $0^*$ &  $-0.7_{-0.3}^{+0.8}$  \\ 
		Norm&   $0.0101_{-0.0010}^{+0.0011}$ &  $0.0101_{-0.0010}^{+0.0011}$ &  $0.0102_{-0.0012}^{+0.0011}$ &  $0.0102_{-0.0012}^{+0.0011}$ \\ \hline
		{\tt simplcutx} \\ 
		$ \Gamma $& $1.621_{-0.012}^{+0.008}$ &   $1.621_{-0.020}^{+0.012}$  &   $1.621_{-0.012}^{+0.009}$ &   $1.621_{-0.020}^{+0.011}$    \\ 
		FracSctr &       $0.370_{-0.021}^{+0.05}$    &  $0.36_{-0.05}^{+0.04}$ &  $0.37_{-0.04}^{+0.05}$ &  $0.37_{-0.05}^{+0.04}$     \\
		ReflFrac &       $1^*$    &  $1^*$ &  $1^*$ &  $1^*$     \\
		$ kT_{\rm e} $  &  $166_{-21}^{+18}$  & $164_{-23}^{+34}$& $166_{-17}^{+23}$ & $167_{-28}^{+36}$  \\ \hline
		{\tt diskbb} \\ 
		$ T_{\rm in} $ [keV]&    $0.86_{-0.19}^{+0.18}$ & $0.86_{-0.07}^{+0.08}$& $0.86_{-0.18}^{+0.4}$ & $0.86_{-0.07}^{+0.08}$   \\ 
		$ p $&       $0.5^*$ &       $0.5^*$ & $0.5^*$ & $0.5^*$    \\ 
		Norm&     $20_{-6}^{+34}$ & $17.1_{-2.3}^{+11}$& $20_{-5}^{+5}$ & $21_{-7}^{+7}$       \\ \hline
		{\tt gauss} \\ 
		$E_{\rm line}$ [keV] &       $6.40^*$  & $6.40^*$ & $6.40^*$  & $6.40^*$  \\ 
		$ \sigma $ [keV]  &     $0.01^*$  & $0.01^*$  & $0.01^*$  & $0.07_{-0.05}^{+0.07}$   \\ 
		Norm/$10^{-3}$& $1.5_{-0.4}^{+0.5}$ & $1.8_{-0.6}^{+0.5}$& $1.6_{-0.4}^{+0.3}$ & $1.6_{-0.5}^{+0.7}$ \\ \hline
		Cross-normalization \\
		$C_{\rm PIN}$ & $1.242_{-0.03}^{+0.021}$ & $1.242_{-0.03}^{+0.021}$ & $1.242_{-0.03}^{+0.022}$ & $1.24\pm0.03$ \\
		$C_{\rm GSO}$ & $1.313_{-0.04}^{+0.020}$ & $1.312_{-0.03}^{+0.020}$ & $1.31\pm0.04$ & $1.31\pm0.04$ \\ \hline
		$\chi^2/\nu $ & $2236.08/2139$ & $2235.81/2138$  & $2235.99/2138$ & $2234.46/2137$  \\ 
		& $=1.04538$ & $=1.04575$ & $=1.04583$ & $=1.04561$ \\
		\hline\hline
	\end{tabular}
	\\
	\caption{Best-fit values from Models~3A and 3B. The reported uncertainties correspond to the 90\% confidence level for one relevant parameter. $^*$ means that the parameter is frozen in the fit. The ionization parameter $\xi$ is in units erg~cm~s$^{-1}$.  
	\label{bestfit_3}}
\end{table*}

\begin{table*}[]
	\centering
	\vspace{0.5cm}
	\begin{tabular}{lcc|cc}
		\hline\hline
		Model            & \multicolumn{2}{c}{4A} & \multicolumn{2}{c}{4B} \\
		& $\alpha_{13}=0$ & $\alpha_{13}$ free & $\alpha_{13}=0$ & $\alpha_{13}$ free \\ 
		\hline
		{\tt tbabs} \\
		$ N_{\rm H}/10^{21}$~cm$^{-2}$ &  $6.0^*$         &  $6.0^* $ &  $6.0^* $        &  $6.0^* $  \\ \hline
		{\tt xstar}   \\ 
		$ N_{\rm H}/10^{21}$~cm$^{-2}$ & $6.4_{-0.6}^{+1.4} $ & $6.4_{-0.6}^{+1.5} $ & $10.6_{-1.5}^{+1.3} $ & $10.6_{-1.3}^{+1.2} $  \\ 
		$\log\xi $&   $3.96_{-0.4}^{+0.13}$  &   $3.96_{-0.4}^{+0.14}$   &   $3.58_{-0.07}^{+0.19}$ &   $3.58_{-0.08}^{+0.17}$  \\ 
		$ z $& $0^* $ & $0^*$  & $0^* $ & $0^*$ \\ \hline
		{\tt relxill\_nk} \\ 
		$ q_{\rm in} $&   $3.7\pm0.4$   &   $3.7\pm0.4$ &   $6.1_{-0.5}^{+0.3}$  &   $6.0_{-0.6}^{+0.4}$      \\ 
		$ q_{\rm out} $&   $3^*$ &   $3^*$ &   $1.25_{-0.21}^{+0.4}$ &   $1.27_{-0.19}^{+0.3}$ \\ 
		$ R_{\rm br} $ [$r_{\rm g}$] &    $>4.9$ &   $>4.9$  &   $10.0_{-3}^{+1.1}$ &   $9.9_{-0.7}^{+3}$ \\ 
		$ a_* $&     $0.954\pm0.011$ &   $0.984_{-0.04}^{+0.006}$  &$0.981_{-0.007}^{+0.003}$&$0.978_{-0.014}^{+0.009}$ \\ 
		$ i $ [deg]&     $43.7_{-1.2}^{+1.0}$ & $43.7_{-1.0}^{+0.9}$& $55.1_{-1.7}^{+1.0}$ & $55.2_{-1.5}^{+1.9}$ \\ 
		$\log\xi $&     $4.22_{-0.07}^{+0.09}$ & $4.21_{-0.07}^{+0.09}$& $4.00_{-0.05}^{+0.03}$ & $4.00_{-0.04}^{+0.05}$ \\ 
		$ A_{\rm Fe} $&     $3.7_{-0.3}^{+0.4}$ & $4.0_{-0.4}^{+0.3}$ & $4.73_{-0.6}^{+0.23}$ & $4.72_{-0.5}^{+0.24}$ \\ 
		$ \alpha_{13} $&       $0^*$    &  $0.20_{-0.9}^{+0.15}$ &  $0^*$ &  $-0.1_{-0.4}^{+0.3}$  \\ 
		Norm&   $0.0352_{-0.0022}^{+0.0023}$ &  $0.0352_{-0.0022}^{+0.0023}$ &  $0.044_{-0.008}^{+0.007}$ &  $0.044_{-0.004}^{+0.0024}$ \\ \hline
		{\tt cutoffpl} \\ 
		$ \Gamma $& $2.66_{-0.04}^{+0.03}$ &   $2.66_{-0.04}^{+0.03}$  &   $2.68_{-0.024}^{+0.05}$ &   $2.68_{-0.03}^{+0.04}$    \\ 
		$ E_{\rm cut} $  [keV] &  $178_{-19}^{+21}$  & $175_{-36}^{+24}$& $144_{-7}^{+37}$ & $145_{-8}^{+25}$  \\ 
		Norm&      $6.3_{-0.3}^{+0.3}$ &  $6.3_{-0.3}^{+0.3}$ &  $7.10_{-0.21}^{+0.4}$ &  $7.11_{-0.21}^{+0.21}$ \\ \hline
		{\tt diskbb}  \\ 
		$ T_{\rm in} $ [keV] &    $0.511_{-0.013}^{+0.012}$ & $0.511_{-0.013}^{+0.013}$& $0.493_{-0.011}^{+0.013}$ & $0.493_{-0.011}^{+0.013}$   \\  
		Norm&     $31168_{-2177}^{+2269}$ & $31382_{-2268}^{+2625}$& $40250_{-2475}^{+1815}$ & $40363_{-2719}^{+2590}$       \\ \hline
		{\tt gauss} \\ 
		$E_{\rm line}$ [keV] &       $6.40^*$  & $6.40^*$ & $6.40^*$  & $6.40^*$  \\ 
		$ \sigma $ [keV] &     $0.01^*$  & $0.01^*$  & $0.01^*$  & $0.01^*$   \\ 
		Norm$10^{-3}$& $0.35_{-0.20}^{+0.20}$ & $0.53_{-0.20}^{+0.20}$& $0.34_{-0.20}^{+0.28}$ & $0.34_{-0.19}^{+0.28}$ \\ \hline
		Cross-normalization\\
		$C_{\rm FPMB}$ & $1.0023_{-0.0013}^{+0.0012}$ & $1.0030_{-0.0014}^{+0.0011}$ & $1.0023_{-0.0013}^{+0.0012}$ & $1.0030_{-0.0013}^{+0.0012}$ \\ \hline
		$\chi^2/\nu $ & $2231.56/1964$ & $2231.33/1963$ & $2209.53/1963$ & $2209.09/1962$ \\ 
		& $=1.13623$ & $=1.13670$ & $=1.12574$ & $=1.12594$ \\
		\hline\hline
	\end{tabular}
     \\
     \caption{Best-fit values from Models~4A and 4B (\textsl{NuSTAR} data of epoch~2). The reported uncertainties correspond to the 90\% confidence level for one relevant parameter. $^*$ means that the parameter is frozen in the fit. The ionization parameter $\xi$ is in units erg~cm~s$^{-1}$. 
     \label{bestfit_4}}
\end{table*}

\begin{table*}[]
	\centering
	\vspace{0.5cm}
	\begin{tabular}{lcc|cc}
		\hline\hline
		Model            & \multicolumn{2}{c}{5A} & \multicolumn{2}{c}{5B} \\
		& $\alpha_{13}=0$ & $\alpha_{13}$ free & $\alpha_{13}=0$ & $\alpha_{13}$ free \\ 
		\hline
		{\tt tbabs} \\
		$ N_{\rm H}/10^{21}$~cm$^{-2}$ &  $6.0^* $        &  $6.0^* $ &  $6.0^* $        &  $6.0^* $  \\ \hline
		{\tt xstar} \\ 
		$ N_{\rm H}/10^{21}$~cm$^{-2}$ & $8.0\pm1.9 $ & $8.2_{-1.6}^{+1.9} $ & $8.4_{-1.6}^{+2.2} $ & $8.3_{-1.6}^{+2.3} $   \\ 
		$\log\xi $ &   $3.90_{-0.022}^{+0.05}$ &   $3.90_{-0.022}^{+0.05}$&   $3.90_{-0.022}^{+0.05}$  &   $3.90_{-0.022}^{+0.06}$   \\ 
		$ z $& $0^* $ & $0^*$  & $0^* $ & $0^*$ \\ \hline
		{\tt relxill\_nk} \\ 
		$ q_{\rm in} $ &  $6.4_{-0.6}^{+1.1}$  &   $5.8_{-0.9}^{+1.2}$&   $6.6_{-1.5}^{+1.8}$   &   $5.5_{-0.7}^{+1.3}$       \\ 
		$ q_{\rm out} $ &   $3^*$ &   $3^*$&   $1.7_{-2.6}^{+5}$ &   $2.0_{-2.5}^{+2.4}$ \\ 
		$ R_{\rm br} $ [$r_{\rm g}$] &   $>4.8$ &   $12_{-7}^{+190}$ &    $9_{-4}^{+180}$ &   $>4.6$  \\ 
		$ a_* $ &$0.955_{-0.016}^{+0.005}$&$0.920\pm0.015$ &  $0.959_{-0.017}^{+0.0023}$ &   $0.920_{-0.024}^{+0.03}$  \\ 
		$ i $ [deg] & $56.6_{-1.6}^{+2.2}$ & $56.4_{-2.4}^{+1.9}$ &     $56.9_{-2.1}^{+4}$ & $56.7_{-1.8}^{+4}$ \\ 
		$\log\xi $ & $3.30_{-0.05}^{+0.03}$ & $3.30_{-0.04}^{+0.03}$ &     $3.30_{-0.04}^{+0.02}$ & $3.30_{-0.04}^{+0.02}$ \\ 
		$ A_{\rm Fe} $ & $4.8_{-0.6}^{+1.4}$ & $4.8_{-0.3}^{+1.0}$ &     $4.8_{-0.6}^{+1.0}$ & $4.9_{-0.6}^{+0.8}$  \\ 
		$ \alpha_{13} $ &  $0^*$ &  $-0.7_{-0.4}^{+0.8}$  &    $0^*$    &  $-0.9_{-0.3}^{+1.3}$  \\  
		Norm &  $0.135_{-0.020}^{+0.019}$ &  $0.134_{-0.011}^{+0.021}$ &   $0.135_{-0.020}^{+0.005}$ &  $0.133_{-0.016}^{+0.006}$  \\ \hline
		{\tt cutoffpl} \\ 
		$ \Gamma $ &   $2.80_{-0.014}^{+0.02}$ &   $2.80_{-0.014}^{+0.02}$ & $2.80_{-0.014}^{+0.02}$ &   $2.80_{-0.014}^{+0.02}$   \\ 
		$ E_{\rm cut} $ [keV] & $246_{-85}^{+84}$ & $234_{-36}^{+110}$ &  $240_{-39}^{+81}$  & $241_{-46}^{+86}$ \\ 
		Norm &  $13.3_{-1.1}^{+0.7}$ &  $13.2_{-0.8}^{+0.7}$ &  $13.2_{-1.4}^{+0.8}$ &  $13.3_{-1.1}^{+0.6}$  \\ \hline
		{\tt diskbb} \\ 
		$ T_{\rm in} $ [keV] & $0.552\pm0.013$ & $0.552\pm0.013$ &    $0.552\pm0.013$ & $0.552\pm0.013$   \\  
		Norm & $25395_{-671}^{+662}$ & $25432_{-1209}^{+594}$ &     $25299_{-1867}^{+851}$ & $25345_{-1363}^{+1723}$  \\ \hline
		{\tt gauss} \\ 
		$E_{\rm line}$ [keV] &       $6.40^*$  & $6.40^*$ & $6.40^*$  & $6.40^*$  \\ 
		$ \sigma $ [keV] &     $0.01^*$  & $0.01^*$  & $0.01^*$  & $0.01^*$   \\ 
		Norm$10^{-3}$ & $1.5\pm0.6$ & $1.3\pm0.6$ & $1.3\pm0.6$ & $1.3\pm0.6$ \\ \hline
		Cross-normalization \\
		$C_{\rm XIS1}$ & $0.9134_{-0.0021}^{+0.0024}$ & $0.9134_{-0.0021}^{+0.0024}$ & $0.9134_{-0.0021}^{+0.0024}$ & $0.9134_{-0.0021}^{+0.0024}$  \\
		$C_{\rm PIN}$ & $0.933_{-0.015}^{+0.03}$ & $0.933_{-0.011}^{+0.024}$ & $0.933_{-0.04}^{+0.024}$ & $0.936_{-0.04}^{+0.025}$  \\
		$C_{\rm GSO}$ & $0.96\pm0.04$ & $0.96\pm0.04$  & $0.97\pm0.04$ & $0.97\pm0.04$ \\ \hline
		$\chi^2/\nu $ & $3419.69/3085$ & $3418.86/3084$ & $3419.18/3084$ & $3418.56/3083$ \\ 
		& $=1.10849$ & $=1.10858$& $=1.10868$ & $=1.10884$  \\
		\hline\hline
	\end{tabular}
	\\
	\caption{Best-fit values from Models~5A and 5B (\textsl{Suzaku} data of epoch~2). The reported uncertainties correspond to the 90\% confidence level for one relevant parameter. $^*$ means that the parameter is frozen in the fit. The ionization parameter $\xi$ is in units erg~cm~s$^{-1}$. 
	\label{bestfit_5}}
\end{table*}

\begin{table*}[]
	\centering
	\vspace{0.5cm}
	\begin{tabular}{lcc}
		\hline\hline
		Model            & $\alpha_{13}=0$ & $\alpha_{13}$ free  \\ \hline 		
		{\tt tbabs} \\
		$ N_{\rm H}/10^{21}$~cm$^{-2}$ &  $5.0_{-0.1}^{+0.1}$   &  $5.0_{-0.1}^{+0.1} $  \\ \hline 		
		{\tt xstar} \\ 
		$ N_{\rm H}/10^{21}$~cm$^{-2}$ & $6.9\pm0.8$  & $6.9\pm0.8 $ \\  
		$\log\xi $ &   $3.91_{-0.03}^{+0.06}$ &   $3.91_{-0.03}^{+0.06}$ \\  
		$ z $& $0^* $ & $0^*$  \\ \hline 
		{\tt relxill\_nk} \\ 
		$ q_{\rm in} $ &  $3.51_{-0.12}^{+0.4}$  &   $3.5_{-0.3}^{+0.3}$  \\  
		$ q_{\rm out} $ &   $3^*$ &   $3^*$ \\ 
		$ R_{\rm br} $ [$r_{\rm g}$] &   $>7.54$ &   $>8.20$ \\  
		$ a_* $ &$0.935_{-0.019}^{+0.012}$&$0.935_{-0.05}^{+0.019}$ \\  
		$ i $ [deg] & $42.3_{-0.7}^{+0.7}$ & $42.3_{-0.6}^{+0.8}$ \\  
		$\log\xi $ & $4.30_{-0.06}^{+0.03}$ & $4.30_{-0.05}^{+0.024}$ \\  
		$ A_{\rm Fe} $ & $7.2_{-0.5}^{+0.7}$ & $7.2_{-0.4}^{+0.7}$ \\ 
		$ \alpha_{13} $ &  $0^*$ &  $0.0_{-1.5}^{+0.3}$  \\   
		Norm &  $0.0233_{-0.0013}^{+0.0012}$ &  $0.0233_{-0.0013}^{+0.0012}$ \\ \hline 
		{\tt cutoffpl} \\ 
		$ \Gamma $ &   $2.55_{-0.03}^{+0.014}$ &   $2.54_{-0.03}^{+0.014}$ \\  
		$ E_{\rm cut} $ [keV]& $93_{-4}^{+3}$ & $92.7_{-1.3}^{+3}$  \\  
		Norm &  $5.38_{-0.24}^{+0.16}$ &  $5.38_{-0.12}^{+0.15}$  \\ \hline
		{\tt diskbb} \\ 
		$ T_{\rm in} $ [keV] & $0.532_{-0.012}^{+0.014}$ & $0.532_{-0.012}^{+0.014}$  \\ 
		Norm & $25818_{-680}^{+841}$ & $25818_{-354}^{+884}$  \\ \hline
		{\tt gauss} \\ 
		$E_{\rm line}$ [keV] &       $6.40^*$  & $6.40^*$ \\ 
		$ \sigma $ [keV]&     $0.01^*$  & $0.01^*$  \\ 
		Norm$10^{-3}$ & $0.52_{-0.15}^{+0.19}$ & $0.52_{-0.15}^{+0.19}$ \\ \hline
		Cross-normalization \\
		$C_{\rm FPMB}$ & $1.0034_{-0.0013}^{+0.0012}$ & $1.0034_{-0.0013}^{+0.0012}$ \\ 
		$C_{\rm XIS0}$ & $1.258\pm0.003$ & $1.258\pm0.003$ \\ 
		$C_{\rm XIS1}$ & $1.148_{-0.011}^{+0.024}$ & $1.148_{-0.011}^{+0.024}$ \\ 
		$C_{\rm PIN}$ & $1.425\pm0.006$ & $1.425_{-0.006}^{+0.007}$  \\ 
		$C_{\rm GSO}$ & $1.54\pm0.13$ & $1.54\pm0.13$ \\ \hline 
		$\chi^2/\nu $ & $5961.42/5024$ & $5961.42/5024$ \\ 
		& $=1.18659$ & $=1.18682$ \\ 
		\hline\hline
	\end{tabular}
	\\
	\caption{Best-fit values from Model~6 (\textsl{NuSTAR}+\textsl{Suzaku} data of epoch~2) The reported uncertainties correspond to the 90\% confidence level for one relevant parameter. $^*$ means that the parameter is frozen in the fit. The ionization parameter $\xi$ is in units erg~cm~s$^{-1}$. \label{bestfit_6}}
\end{table*}

\section{Discussion and conclusions \label{s-d-c}}

Generally speaking, all our fits (either for $\alpha_{13} = 0$ or for $\alpha_{13}$ free) are consistent with previous studies and we find a spin parameter close to 1 and an inclination angle around $40^\circ$~\cite{c2,c1,c3,c4,c6,c7,c8}. While we have always modeled the emissivity profile of the disk with a broken power-law and assumed either $q_{\rm out} = 3$ or free, the final results are quite independent of this last choice.

For the 2009 \textsl{Suzaku} observation with the source in the hard state, we have six models (1A, 1B, 2A, 2B, 3A, 3B). Model~1 provides the worse fit and we do not recover the Kerr metric at 90\% confidence level (see Tab.~\ref{bestfit_1}). Model~2 improves the fit ($\Delta\chi^2$ is around 80 without adding new free parameters) and we recover the Kerr spacetime (see Tab.~\ref{bestfit_2}). The spin parameter and the inclination angle of the disk do not change much between Model~1 and 2, while we can significantly reduce the iron abundance. We note, however, that in Model~2 we get quite a high value of $T_{\rm in}$ in {\tt diskbb}, too high for a source in the hard state. Model~3 improves the fit of Model~2 decreasing $\chi^2$ by about 60 and keeping the same number of free parameters. In Model~3, we recover the Kerr metric at 90\% confidence level, we decrease the value of $T_{\rm in}$ in {\tt diskbb}, the iron abundance increases a bit with respect to Model~2 but it is still lower than Model~1, and the estimates of the black hole spin parameter and of the inclination angle of the disk do not change much.

For the observations of the source in the soft state in 2012, we see that there is more difference between the models with $q_{\rm out} = 3$ and those with $q_{\rm out}$ free. The emissivity profile of the disk is determined by the coronal geometry, which can vary with a timescales of a few days, and therefore it is not a surprise if the observations in 2009 and 2012 present some differences in the emissivity profile. Moreover, one is in the hard state and the other observation is in the soft state, so the coronal geometry could be, in principle, very different. The \textsl{NuSTAR} fits (Model~4) and the \textsl{Suzaku} fits (Model~5) are, in general, consistent, but we note that the constraints on the deformation parameter $\alpha_{13}$ are better in the \textsl{NuSTAR} fits, suggesting that the key-point to get stringent tests of the Kerr metric is not to have a good energy resolution near the iron line, or at least that this is not enough. We also note that, with the exception of Model~4, the estimate of the inclination angle of the disk is around $55^\circ$, which is a bit higher than what we would expect. When we combine the \textsl{NuSTAR} and \textsl{Suzaku} spectra, the estimate of the inclination angle of the disk returns to a value around $40^\circ$.

Unlike in the analysis of the \textsl{NuSTAR} observations of Cygnus~X-1 presented in Ref.~\cite{Liu:2019vqh}, in the present work we are able to constrain the deformation parameter $\alpha_{13}$ of the Johannsen metric. In the observation in the hard state, with Model~3A we find (90\% CL)
\be
\alpha_{13} = -0.2_{-0.7}^{+0.3} \, .
\ee
In the observation in the soft state, when we fit the \textsl{NuSTAR} and \textsl{Suzaku} spectra (Model~6), we find (90\% CL)
\be
\alpha_{13} = 0.0_{-1.5}^{+0.3} \, .
\ee
These constraints are not among the best ones using {\tt relxill\_nk}. For example, limiting the discussion to tests of the Kerr metric with stellar-mass black holes, from the 2007 \textsl{Suzaku} observation of GRS~1915+105, we got~\cite{Abdikamalov:2020oci} (90\% CL)
\be
\alpha_{13} = 0.00_{-0.15}^{+0.05} \, .
\ee
From the simultaneous analysis of reflection features and thermal spectrum of GX~339--4 with \textsl{Swift} and \textsl{NuSTAR} data, with a broken power-law emissivity profile we found~\cite{Tripathi:2020dni} (90\% CL)
\be
\alpha_{13} = -0.010^{+0.024}_{-0.018} \, .
\ee

Cygnus~X-1 thus remains quite a complicated source, especially because it is a high-mass X-ray binary and therefore the radiation emitted from the very inner region of the accretion disk is inevitably absorbed by the strong wind from the companion star. However, here we have definitively improved our analysis of Ref.~\cite{Liu:2019vqh}, where it was not possible to constrain the deformation parameter $\alpha_{13}$.

We remind that all uncertainties reported in this manuscript only refer to the statistical uncertainties, while the systematic ones are ignored. Systematic uncertainties mainly include simplifications in the theoretical model employed to analyze the data, but there are even instrumental and data analysis uncertainties; for a review, see \cite{Bambi:2020jpe} and references therein. We note that this is not only an issue for testing general relativity, but for any measurement of the properties of accreting black holes using X-ray reflection spectroscopy.

Among the systematic uncertainties related to the theoretical model, simplifications in the description of the structure of the accretion disk are often thought to be among the most important source of uncertainty. Relativistic reflection models normally employ the Novikov-Thorne model for the description of the accretion disk~\cite{ntm73,Page:1974he}. Such a model is valid for geometrically thin and optically thick accretion disks. If we employ the model to analyze sources with thick accretion disks, the final measurements can be affected by large systematic errors. If we assume general relativity and we want to measure the black hole spin, we can easily get an incorrect spin measurement~\cite{Riaz:2019kat}. In the case of tests of the Kerr metric, we can obtain a non-vanishing measurement of the deformation parameter even if the metric around the compact object is described by the Kerr solution~\cite{Riaz:2019bkv}. Assuming general relativity, numerical simulations show that the Novikov-Thorne model provides a good description of thin accretion disks when the accretion disk luminosity of the source is between a few percent and up to about 30\% of its Eddington limit, and that systematic uncertainties in the estimate of the black hole spin parameter are small; see, e.g., Refs.~\cite{Reynolds:2007rx,Penna:2010hu,Kulkarni:2011cy}. However, numerical simulations with different magnetic field configurations may lead to different conclusions~\cite{Noble:2010mm}.

The take-away message of the available studies is that X-ray reflection spectroscopy measurements employing the Novikov-Thorne model can be accurate if we select the sources and the observations with the right characteristics, while incorrect measurements can be easily obtained when we use the model for any source showing a spectrum with reflection features. We also note that a reflection model is unsuitable to test the validity of the Novikov-Thorne model, in the sense that we can obtain a good fit even if the disk is not thin because of parameter degeneracy~\cite{Riaz:2019kat,Riaz:2019bkv}. For example, even within the Kerr metric, it is difficult to test the Keplerian disk velocity of the material in the accretion disk without an independent measurement of the disk's inclination angle~\cite{Tripathi:2020wfi}. In the observations analyzed in this work, the accretion disk luminosity can be estimated to be around 1\% in epoch~1 and around 10\% in epoch~2. Especially in the second observation, in which the source is in the soft state, we can expect that the systematic uncertainties from the Novikov-Thorne model are modest and do not have a significant impact on our measurements.


\vspace{0.5cm}

{\bf Acknowledgments --}
This work was supported by the Innovation Program of the Shanghai Municipal Education Commission, Grant No.~2019-01-07-00-07-E00035, the National Natural Science Foundation of China (NSFC), Grant No.~11973019, and Fudan University, Grant No.~JIH1512604.


\end{document}